\documentclass[10pt,twocolumn,letterpaper]{article}

\usepackage{wacv}      

\usepackage{graphicx}
\usepackage{amsmath}
\usepackage{amssymb}
\usepackage{booktabs}
\usepackage{multirow}
\usepackage{adjustbox}
\usepackage{rotating}
\usepackage{xcolor}
\usepackage[symbol]{footmisc}

\usepackage[pagebackref,breaklinks,colorlinks]{hyperref}

\usepackage[capitalize]{cleveref}
\crefname{section}{Sec.}{Secs.}
\Crefname{section}{Section}{Sections}
\Crefname{table}{Table}{Tables}
\crefname{table}{Tab.}{Tabs.}

\def\wacvPaperID{1305} 
\def\confName{WACV}
\def\confYear{2025}

\begin{document}

\title{UnDIVE: Generalized Underwater Video Enhancement Using Generative Priors}

\author{Suhas Srinath$^1$ \quad Aditya Chandrasekar$^{1, 2 \dag}$ \quad Hemang Jamadagni$^3$ \quad Rajiv Soundararajan$^1$ \\
Prathosh A P$^1$\\
$^1$ Indian Institute of Science \quad $^2$ Qualcomm \quad $^3$ National Institute of Technology Karnataka 
}
\maketitle

\begin{abstract}
With the rise of marine exploration, underwater imaging has gained significant attention as a research topic. Underwater video enhancement has become crucial for real-time computer vision tasks in marine exploration. However, most existing methods focus on enhancing individual frames and neglect video temporal dynamics, leading to visually poor enhancements. Furthermore, the lack of ground-truth references limits the use of abundant available underwater video data in many applications. To address these issues, we propose a two-stage framework for enhancing underwater videos. The first stage uses a denoising diffusion probabilistic model to learn a generative prior from unlabeled data, capturing robust and descriptive feature representations. In the second stage, this prior is incorporated into a physics-based image formulation for spatial enhancement, while also enforcing temporal consistency between video frames. Our method enables real-time and computationally-efficient processing of high-resolution underwater videos at lower resolutions, and offers efficient enhancement in the presence of diverse water-types. Extensive experiments on four datasets show that our approach generalizes well and outperforms existing enhancement methods. Our code is available at \small \url{github.com/suhas-srinath/undive}.
\end{abstract}

\section{Introduction}
\label{sec:intro}

The goal of underwater enhancement is to reduce artifacts and recover lost colors from the water scattering effect \cite{scatter} in images and videos. 
Underwater enhancement finds applications in areas such as coral reef monitoring \cite{coral1}, archaeology \cite{archae1} and underwater robotics \cite{robot}. 
Underwater video enhancement (UVE) is often challenging due to multiple reasons. 
Collecting high-quality videos amidst distortions like blur, reduced illumination, complex channel attenuation, and obtaining ground-truth video data to supervise learning-based algorithms are extremely cumbersome.

\begin{table}[!ht]
\centering
\setlength{\tabcolsep}{0.2pt}
\begin{tabular}{cccc}
\begin{turn}{90} \hspace{0.2cm} Input \end{turn} & \hspace{0.1mm}
\includegraphics[width=0.15\textwidth]{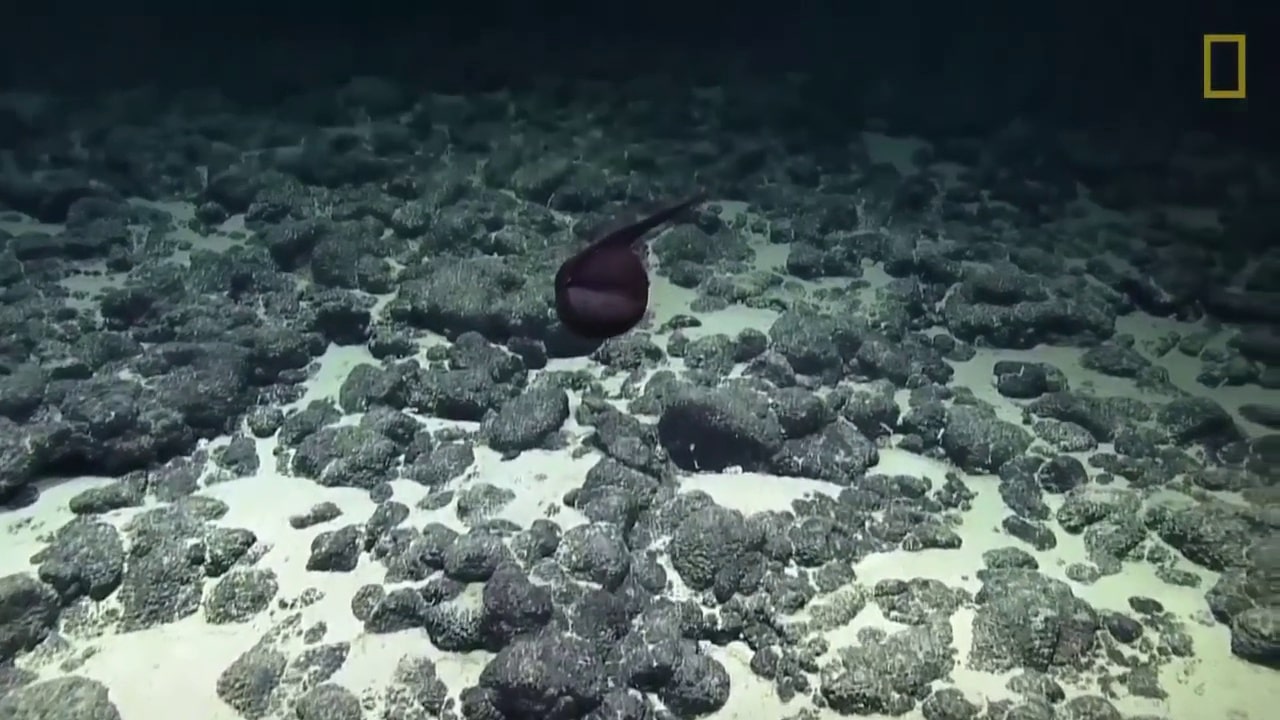}     &    \includegraphics[width=0.15\textwidth]{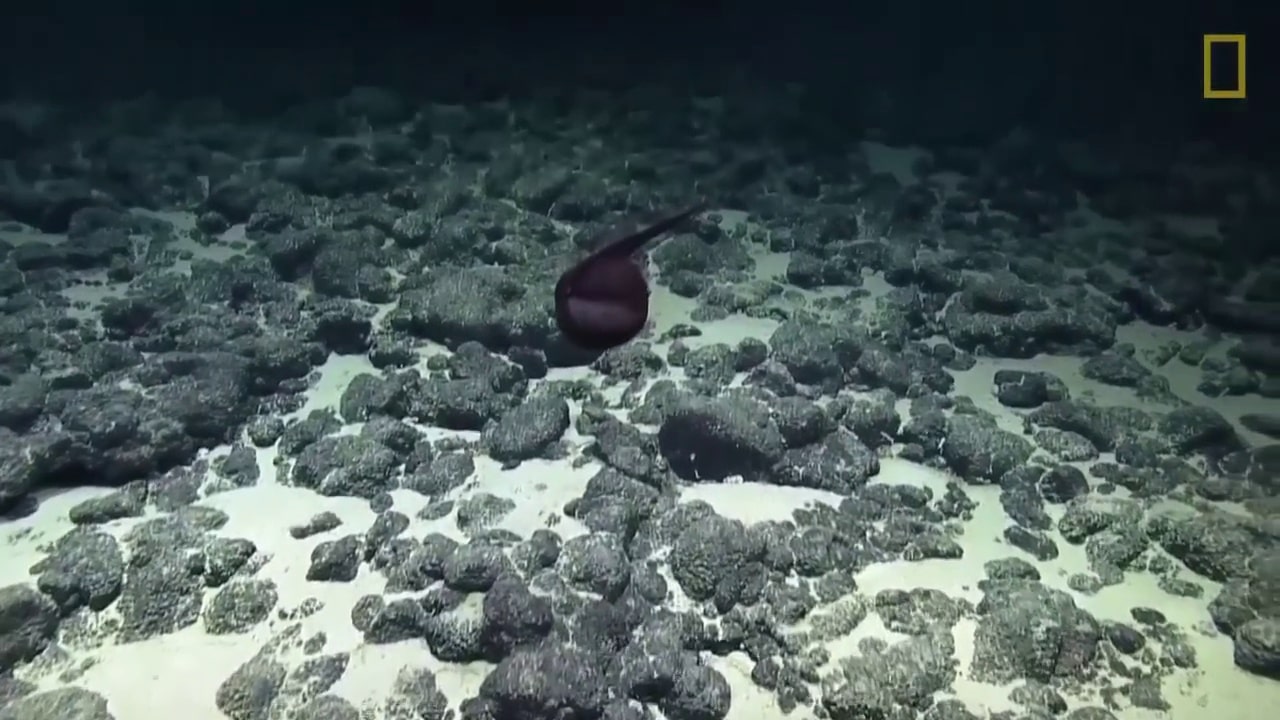} &  \includegraphics[width=0.15\textwidth]{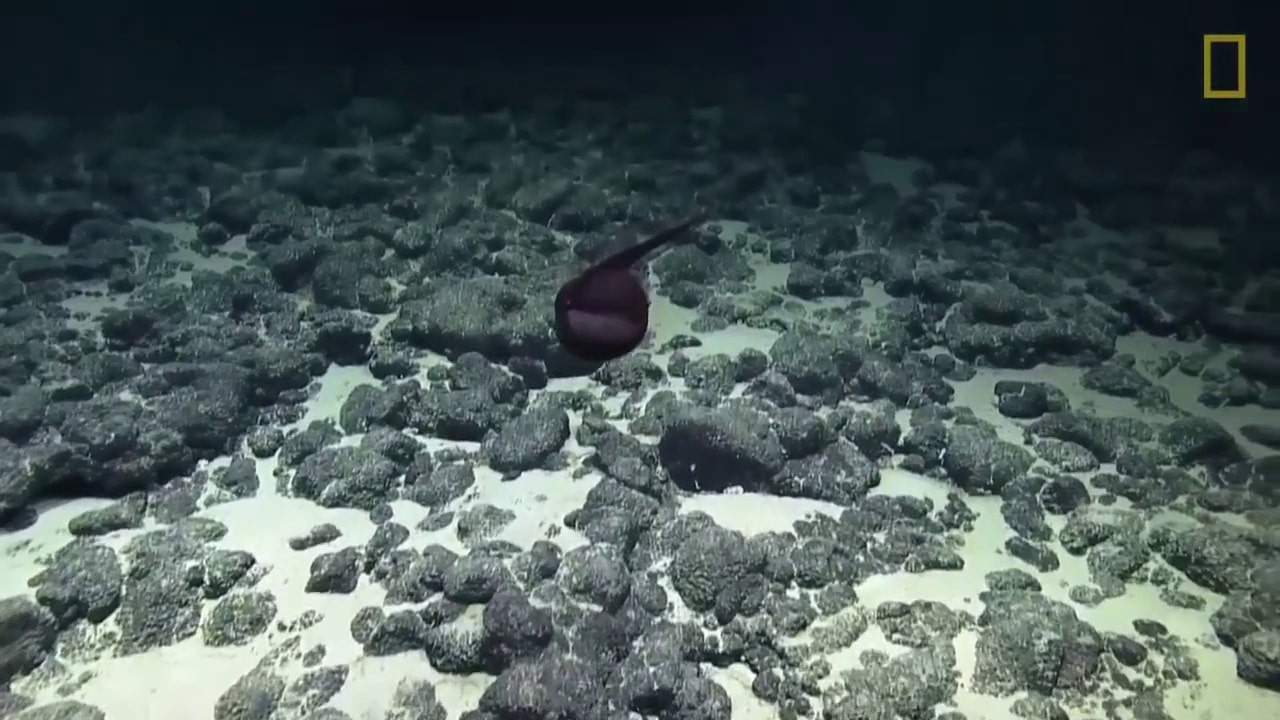}  \\
\begin{turn}{90} PhISHNet \end{turn} & \hspace{0.1mm}
\includegraphics[width=0.15\textwidth]{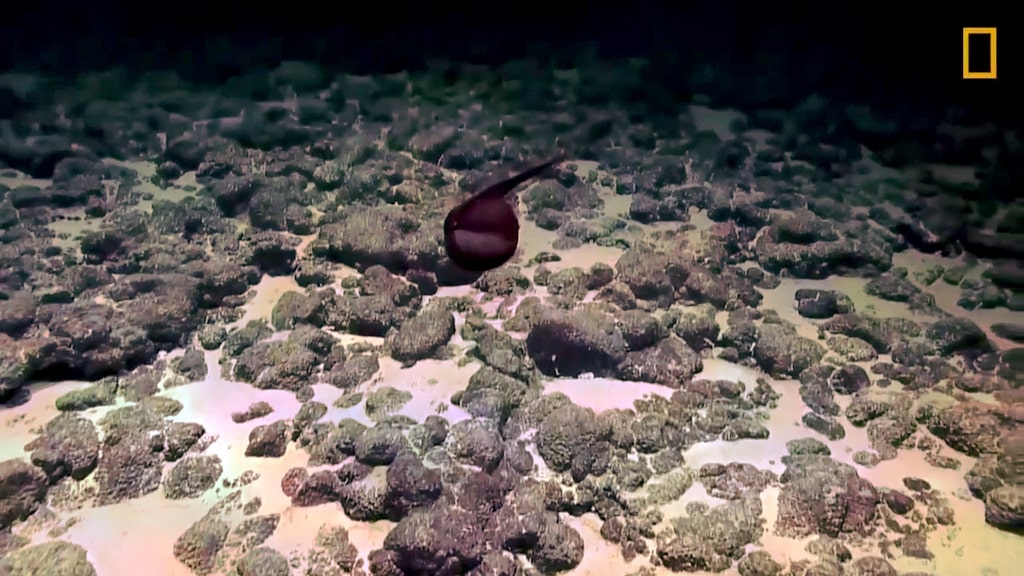}     &    \includegraphics[width=0.15\textwidth]{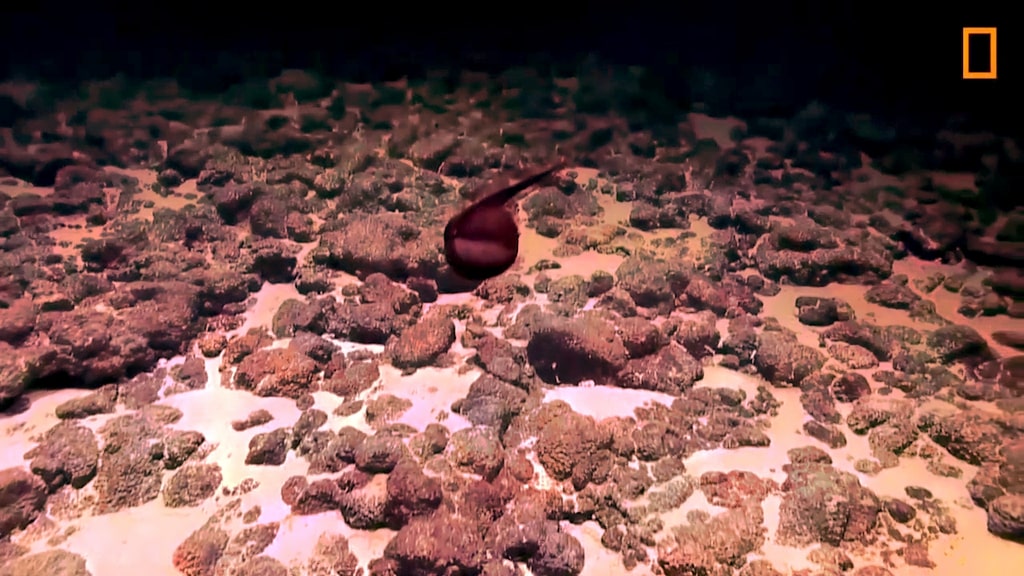} &  \includegraphics[width=0.15\textwidth]{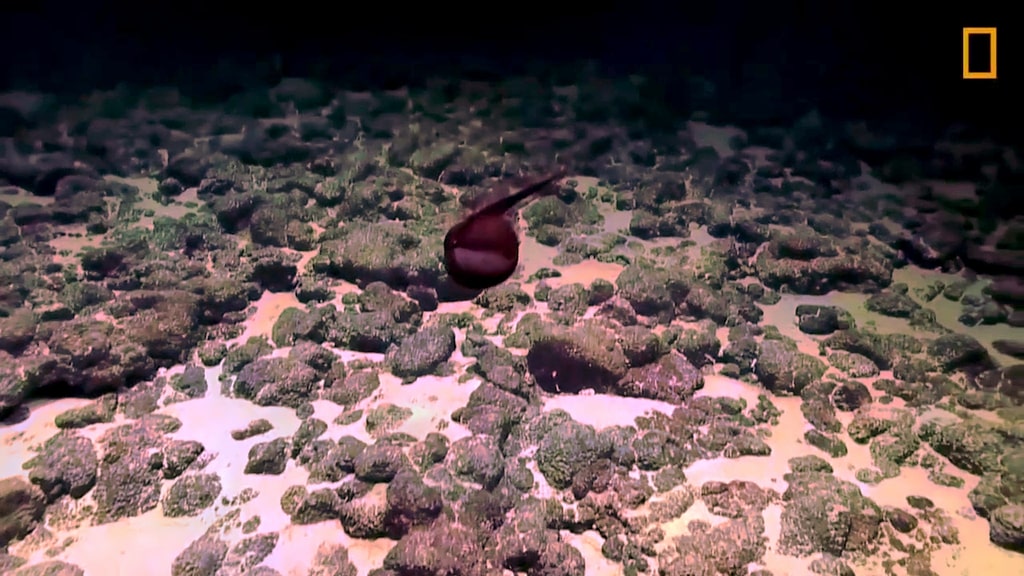}  \\
\begin{turn}{90} \hspace{0.3cm} Ours \end{turn} & \hspace{0.1mm}
\includegraphics[width=0.15\textwidth]{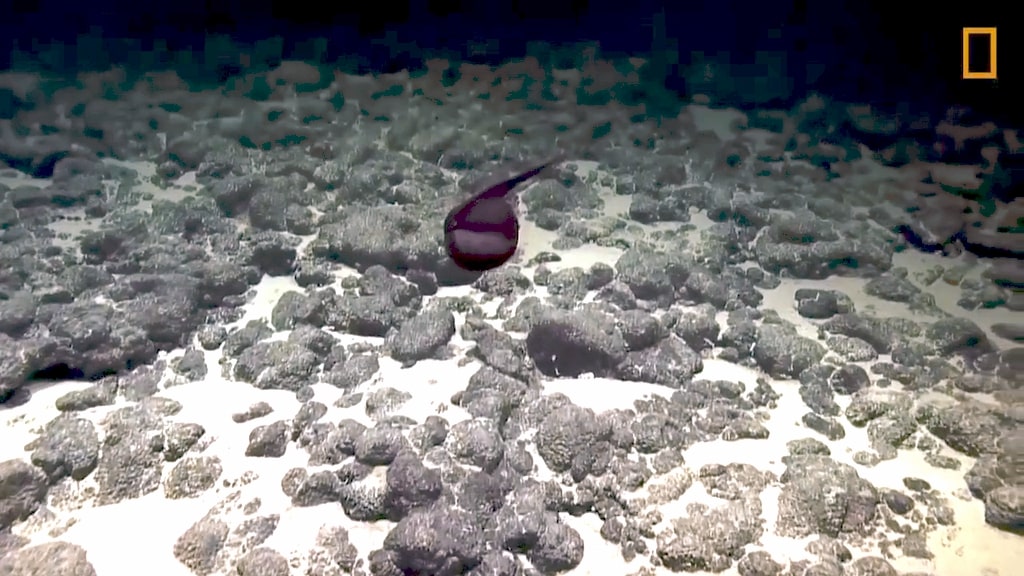}     &    \includegraphics[width=0.15\textwidth]{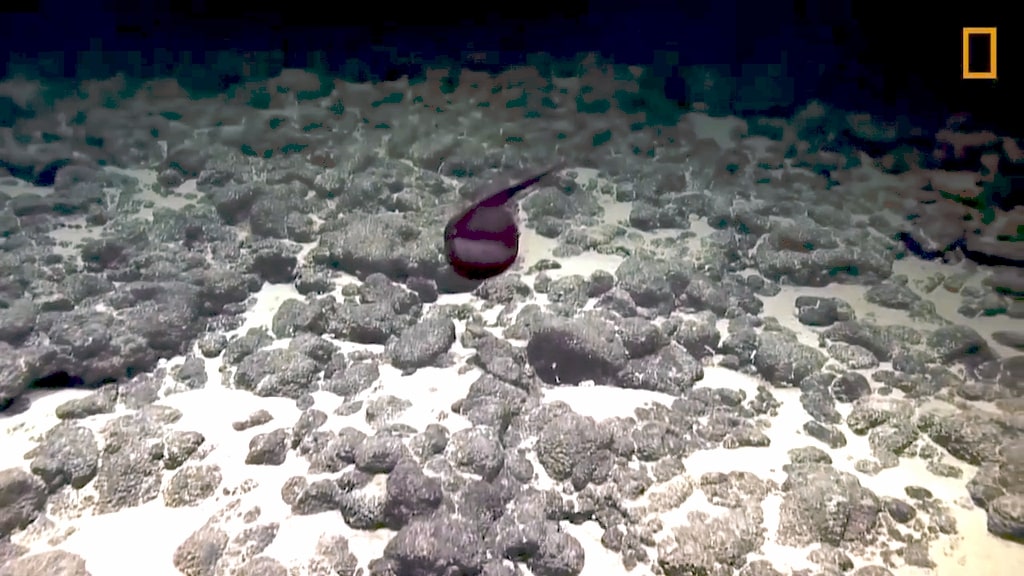}     & 
\includegraphics[width=0.15\textwidth]{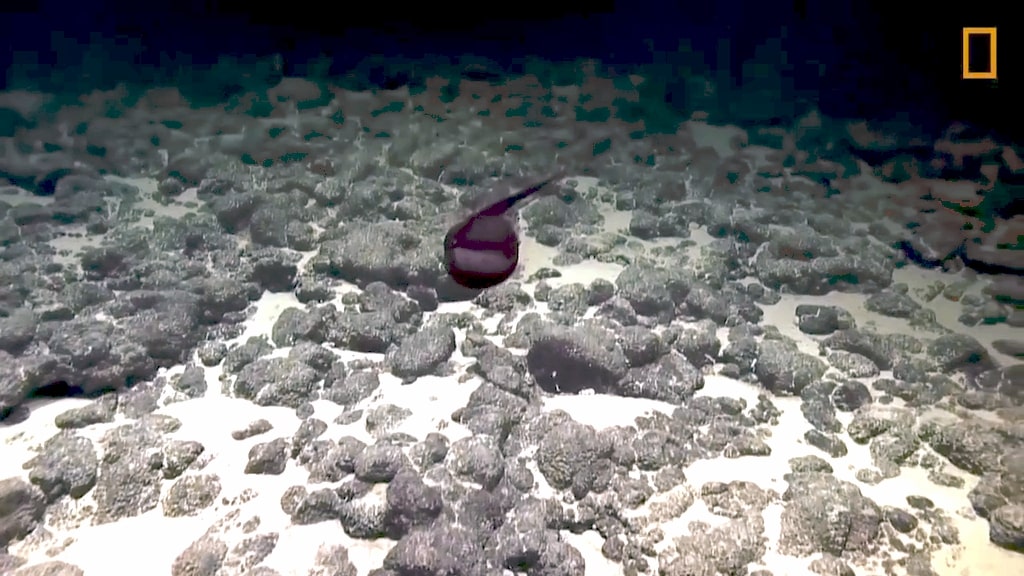} \\
& \hspace{0.1mm} (a) Frame $x_{t-1}$ & (b) Frame $x_{t}$ & (c) Frame $x_{t+1}$ \\
\end{tabular}
\captionof{figure}{UnDIVE (bottom row) enhances contiguous video frames (top row) from the UOT32 \cite{uot32} dataset (DeepSeaFish video), while maintaining consistent colors and illumination as opposed to image-based methods such as PhISH-Net (middle row).}
\label{fig:teaser}
\end{table}

Recent works \cite{uieb,uveb,uve1,uve2,uve3}\footnote[2]{Work done while at Indian Institute of Science.} try to solve UVE by training convolutional neural networks through the supervision of large-scale ground truth data, but often fail to generalize to diverse and unseen water types.
While learning-based methods \cite{uieb,uie1,uie2,uie3} have been reasonably successful in achieving underwater image enhancement (UIE), they cannot be directly scaled to UVE since they do not account for object motion in videos.
Moreover, variations in the illumination across enhanced frames from UIE methods cause flicker-like artifacts in the enhanced videos. 
Despite the greater requirement of videos than images in marine applications, very few UVE methods exist.  

UIE methods have been very successful in restoring the color and contrast in underwater scenes. 
Earlier UIE methods attempted to solve enhancement through pixel adjustments and classical priors \cite{uieclassical1, uieclassical2}. 
With the development of deep learning, data-centric methods \cite{uiegan1,uiegan2,uieb} leveraged paired image training to learn good enhancement. 
To generalize better, unsupervised methods \cite{usuir} have also been developed for UIE.
Despite numerous advancements in UIE, scaling these techniques to videos via frame aggregation remains challenging due to the lack of temporal alignment.

To address the aforementioned challenges, we propose to solve the UVE problem through the introduction of temporal consistency into a physics-driven spatial enhancement network that mitigates artifacts that arise due to water scattering. 
Firstly, to learn efficient spatial enhancement, we propose to learn a generative prior through a self-supervised denoising diffusion probabilistic model (DDPM) \cite{ddpm} that learns robust representations of underwater images.
To the best of our knowledge, this is the first work that learns a generative prior using diffusion for UVE.
The encoder of the UNet learned by the DDPM is subsequently integrated into the UVE framework for efficient downstream enhancement. 

In general, video enhancement \cite{videoenhancement1,videoenhancement2,videoenhancement3} and restoration methods \cite{vrt1,vrt2} incorporate motion information during the learning process to make videos more temporally consistent through the alignment of objects or representations between consecutive frames. 
However, such transformer-based methods are prone to overfitting and do not generalize well (simultaneously perform well on diverse datasets and water-types).
To tackle this issue, we propose to incorporate motion into UVE through an unsupervised optical flow loss so that objects in consecutive enhanced frames exhibit smoother motion and uniform illumination, colors, and contrast. 

Leveraging the generalization capability of the generative prior and the unsupervised temporal consistency loss, we propose an \textbf{Un}derwater \textbf{D}omain \textbf{I}ndependent \textbf{V}ideo \textbf{E}nhancement (\textbf{UnDIVE}) framework that can efficiently process high-resolution videos with fairly low complexity and inference times.  
To summarize, our contributions are as follows:

\begin{itemize}
    \item A two-stage training framework for UVE. The first stage learns a prior on underwater images, and the second stage utilizes this prior to learn spatial and temporal enhancement.
    \item A generative prior trained on carefully chosen underwater images learned through a DDPM to provide robust representations. The learned encoder is subsequently integrated into the UnDIVE network for downstream enhancement.
    \item An unsupervised temporal consistency loss to incorporate motion information into a physics-driven spatial enhancement network, enabling enhanced videos to maintain uniform illumination, accurate colors, and improved contrast.
    \item Through extensive experiments, we demonstrate the superiority and generalizability of UnDIVE over other enhancement methods on four diverse underwater video datasets on multiple no-reference (NR) visual quality metrics.
\end{itemize}

\section{Related Work}
\label{sec:rel_work}

\subsection{Underwater Image/Video Enhancement} 

UIE methods can be broadly categorized into traditional (image-processing or model-based) and learning-based approaches. 
Traditional methods aim to restore degraded underwater images using prior visual characteristics or by treating UIE as an inversion problem, reversing the degradation caused by the imaging process.
Popular methods employ various image processing techniques \cite{ip1,ip2,ip3,ip4,ip5,ip6,ip7,ip8}, utilize priors \cite{prior1,prior2,prior3,prior4,prior5,prior6,prior7,prior8,prior9}, or rely on an underwater image formation model \cite{imform1,imform2,imform3,imform4,imform5,imform6,imform7}.
Learning-based methods, particularly deep learning methods, have become prominent in producing high-quality results. 

Although some methods do not account for the physical process of underwater image formation, they learn to reverse the degradation using large-scale training. 
Notable CNN-based methods include WaterNet \cite{uieb}, UWCNN \cite{uwcnn}, SGUIE-Net \cite{sguie}, UICoE Net \cite{uve38k}, LANet \cite{lanet}, PUIE-Net \cite{puienet}, Semi-UIE \cite{semiuie}, PhISH-Net \cite{PhISHnet}, UIE-Net \cite{uienet}, USUIR \cite{usuir}, and URanker \cite{uranker}.
While these methods are successful, they generally rely on large-scale training with ground-truth references, which is difficult and time-consuming to obtain.  
In the context of UVE, most of these works apply UIE on a frame-by-frame basis, often resulting in visual artifacts like inconsistent lighting across frames and flicker.
We propose learning spatial enhancement along with temporal consistency to mitigate these issues.

\begin{figure*}[!t]
    \centering
    \includegraphics[width=\textwidth]{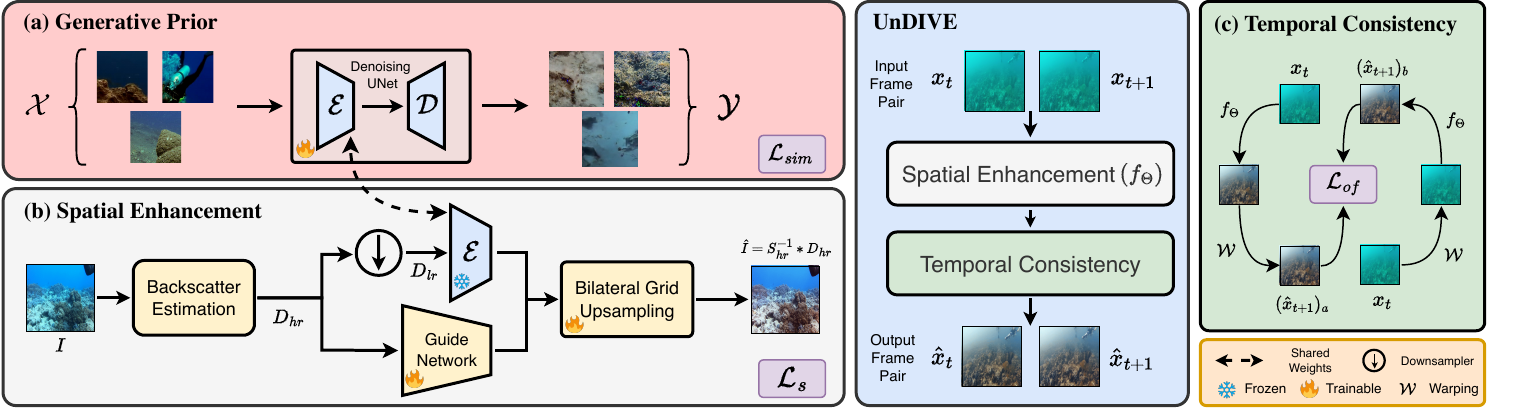}
    \caption{Overall framework of UnDIVE. (a) The first stage learns a generative prior on underwater images, where a denoising DDPM UNet is trained with the loss $\mathcal{L}_{sim}$. (b) The second stage utilizes the trained encoder, and learns the spatial enhancement ($f_{\Theta}$) with loss $\mathcal{L}_s$. First, backscatter is removed, and the image $D_{hr}$ is processed through a guide network to capture low-level local details, while the downsampled (by two) image is passed through $\mathcal{E}$ capturing global (high-level) information. Finally, both streams are fused and upsampled to match the input resolution. (c) A temporal consistency loss $\mathcal{L}_t$ enforces uniform illumination and colors in the enhanced frames.}
    \label{fig:undive}
\end{figure*}

\subsection{Generative Models for Enhancement}

With the success of generative adversarial networks (GANs), various GAN-based methods have been developed, enabling faster computation.
GAN-based approaches include UGAN \cite{uiegan2}, UW-GAN \cite{uwgan}, Spiral-GAN \cite{spiralgan}, FunIE-GAN \cite{funiegan}, WaterGAN \cite{uiegan1}, Dense GAN \cite{densegan}, FEGAN \cite{fegan}, TOPAL \cite{topal}, CycleGAN \cite{cyclegan} and CLUIE-Net \cite{cluienet}. 
PUIE-Net \cite{puienet} uses a conditional variational autoencoder to generate an enhancement distribution.

Although generative models are shown to run with excellent computational speeds, they often trade-off performance due to undesired hallucinated colors. 
Such methods are unable to leverage the generalizability of generative models, and to this end, we propose to use a generative prior through a DDPM that learns good representations of underwater scenes.
This prior enables efficient transfer learning for the downstream UVE task.

\subsection{Temporal Consistency in Videos}

Flickering tends to occur when single-frame-based methods are applied to video clips, leading to notable visual incoherence. 
Efforts to use image-based methods to enhance videos or enforce ``temporal consistency'' have been explored in various contexts, such as video-to-video synthesis \cite{temp_consistency_1}, video enhancement \cite{tempconst_sing}, style-transfer \cite{temp_consistency_2}, depth estimation \cite{tempconst_video} and semantic segmentation \cite{temp_consistency_3}, to mitigate flickering issues.

These methods typically employ self-consistency by enforcing the similarity of data pairs \cite{tempsing1,tempsing2} or explicitly learn temporal consistency using additional modules \cite{tempsing3} to improve the performance and stability of deep models. 
For instance, Zhang \textit{et al.} \cite{tempconst_sing} implicitly embed temporal consistency using optical flow \cite{hornschunkof} generated from single images. 
In this work, we leverage both forward and backward optical flow to learn visual consistency between a pair of consecutive frames of a video. 
This allows the model to learn consistent reconstructions of color, structure, and illumination in both directions to enable uniform enhancement.

\section{UnDIVE}
\label{sec:method}

Our framework, UnDIVE, consists of learning a generative prior on underwater images, incorporating it into a spatial enhancement network and further introducing temporal consistency into the enhancements.
An overview of the training framework is illustrated in Fig. \ref{fig:undive}.

\subsection{Generative Prior for Underwater Images}

DDPMs \cite{ddpm} are generative models that have shown tremendous success in various applications for their ability to learn generalizable feature encoders using unlabeled data.
These encoders can then be downstreamed effectively for multiple transfer learning tasks.
In our work, we leverage such an encoder to learn robust representations on underwater images.
The training images are chosen such that the intensity histograms are closer to uniform to ensure the presence of significant objects.
Since enhancement and generation are image-to-image tasks, we find it appropriate to train a UNet-based DDPM for generation and transfer it for the task of enhancement.
The following paragraphs describe the training procedure of the DDPM.

Consider a UNet characterized by an encoder $\mathcal{E}$ and a decoder $\mathcal{D}$ (refer Fig. \ref{fig:undive}) parameterized by $\theta$. 
Let $\mathbf{x}_0\in \mathcal{X}$ be an input image to $\mathcal{E}$, and $\mathbf{x}_t$ be the corresponding noisy image at time step $t=1,2, \dots, T$. 
$\mathbf{x}_t$ is obtained from $\mathbf{x}_{t-1}$ according to the diffusion process:
\begin{equation}\label{eq2}
    \mathbf{x}_t = \sqrt{1-\beta_t}\mathbf{x}_{t-1} + \sqrt{\beta_t}\epsilon_t,
\end{equation}
where $\beta_t$ is the noise schedule parameter at time $t$, and $\epsilon_t \sim \mathcal{N}(\mathbf{0},\mathbf{I})$,  $\forall t=1,2, \dots, T$. 
The distributions of the forward diffusion process are denoted by  $q(\mathbf{x}_t|\mathbf{x}_{t-1})$, which are assumed to follow a first-order Gaussian Markov process. 
The reverse (decoding process) is modeled using a parametric family of distributions denoted by $p_\theta(\mathbf{x}_{t-1}|\mathbf{x}_{t})$. 

The formulation in \eqref{eq2} allows us to sample $\mathbf{x}_t$ directly from the original input image $\mathbf{x}_0$ as $\mathbf{x}_t = \sqrt{\Bar{\alpha}_t}\mathbf{x}_{0} + \sqrt{1-\Bar{\alpha}_t}\epsilon$,
where $\Bar{\alpha}_t=\prod_{i=1}^{t} \alpha_i$ and $\alpha_t=1-\beta_t$. If the noise schedule parameters $(\beta_t)_{t=1}^{T}$ are very small such that $\beta_T \rightarrow 0$, then the distribution of $\mathbf{x}_T$ can be well approximated by the standard Gaussian distribution i.e., $q(\mathbf{x}_T) \sim \mathcal{N}(\mathbf{0},\mathbf{I})$. 
The goal of the DDPM is to estimate the parameters of $p_\theta$ 
by optimizing a variational lower bound on the log-likelihood of the data $\mathbf{x}_0$ under the model $p_\theta$. 
A simplified loss function used to train the DDPM is as follows:
\begin{equation}\label{eq3}
    \mathcal{L}_{\text{sim}} = \sum_{t \geq 1} L_{t} \text{ , where }
    L_{t} = \underset{\mathbf{x}_{0}, \epsilon}{{\mathbb{E}}} \Bigr[\| \epsilon - \epsilon_{\boldsymbol{\theta}}(\mathbf{x}_{t}, t) \|^{2}\Bigr],
\end{equation}
where $\epsilon$ and $\epsilon_\theta(\mathbf{x}_t, t)$ correspond to the input (real) noise and the predicted noise at time step $t$, respectively. Once the DDPM is trained, the encoder $\mathcal{E}$ of the UNet at time step $t=0$ (since there is no requirement of any noisy versions of $\mathbf{x}_0$) is used ($\mathcal{E}$ is kept frozen) in the subsequent spatial enhancement stage to provide robust representations.

\subsection{Learning Spatial Enhancement}

Since the colors and contrast in underwater scenes are lost due to the water scattering effect, it is necessary to mitigate this to recover objects present in the scene.
The scattering effect towards the surface causes noisy artifacts, called backscatter, that proportionally increases with depth \cite{seathru, PhISHnet}.\\

\noindent \textbf{Backscatter Estimation:} A simple linear formulation \cite{seathru, widebandcoeff} that models the backscatter was proposed as follows:
\begin{equation}
    I_c = D_c + B_c,
    \label{eq:backscatter}
\end{equation}
where $I_c$ is the c$^{th}$ channel of an RGB image $I$, $c \in \{ r,g,b\}$, $D_c$ is the direct signal and $B_c$ is the backscatter.
$B_c$ can be estimated as a function of wideband attenuation coefficients and the depth map. 
Since knowledge of depth is required for estimating the backscatter, we employ the SlowTV monocular depth estimator \cite{slowtv1, slowtv2} to generate robust depth maps in a computationally efficient manner.
SlowTV, trained on a large dataset of images, including underwater scenes, is well-suited for depth estimation. Removing backscatter eliminates water from the input image, making the model independent of water-type and enhancing generalization..\\ 

\noindent \textbf{Training Spatial Enhancement:} Once the backscatter is removed from the input image $I$, we utilize the backbone of PhISH-Net \cite{PhISHnet} to obtain a high resolution illumination map $S_{hr}$. 
The key difference from PhISHNet is in the low-resolution feature encoder. We then use the encoder $\mathcal{E}$ to enhance the estimation of the illumination map by improving the structural representations of underwater scenes. The guide network captures low-level local features, which are fused with the high-level features from $\mathcal{E}$ resulting in a complete representation. The fused features are then bilinearly upsampled (with learnable weights) to match the input resolution.
The final enhanced image $\hat{I}$ can be estimated from the high-resolution image $I_{hr}$ as:
\begin{equation}
    \hat{I} = \frac{I_{hr}}{S_{hr} + \epsilon},
    \label{eq:PhISHrecon}
\end{equation}
a pixel-wise division where $\epsilon > 0$ is a constant that ensures numerical stability. 
We utilize three spatial losses to train UnDIVE on a set of paired underwater images $\{ (\hat{I}, I^{gt})\}$ ($I^{gt}$ is the corresponding ground truth for $I$).\\

\noindent \textbf{Reconstruction loss:} To reconstruct the structure, color, illumination, and sharpness in the enhanced image, we employ a loss function consisting of a combination of structural similarity measure (SSIM) and the L$_1$ distance at a pixel level.
The loss is expressed as 
\begin{equation}
    \mathcal{L}_r = 0.85 (1 - \text{SSIM}(\hat{I}, I_{gt})) + 0.15 \sum_j | \hat{I}_j - I^{gt}_j |, 
    \label{eq:loss_rec}
\end{equation}
where $j$ corresponds to the pixel locations in each image.
This loss function has shown tremendous success in learning unsupervised optical flow \cite{arflow} and in multiple view synthesis applications.\\

\noindent \textbf{Smoothness Loss:} To maintain uniform illumination in the enhancements, we impose a smoothness loss $\mathcal{L}_{sm}$ which is a sum of weighted L$_2$ norm of the gradients of the illumination map $S_{hr}$ as follows:
\begin{equation}
    \mathcal{L}_{sm} = \sum_j w_j \| \nabla_j S_{hr} \|_2,
    \label{eq:loss_smooth}
\end{equation}
where $j$ denotes the pixel location and $w_j$ is a weight corresponding to each spatial location. \\

\noindent \textbf{Color Loss:} Additionally, a color loss is introduced to keep the reconstructed intensities consistent with that of the ground truth image. 
The loss is expressed as the angle between the pixel intensities vectorized across the channels of the input and ground truth image as
\begin{equation}
    \mathcal{L}_c = \sum_{j} \text{arccos} \frac{\hat{I}_j \cdot I^{gt}_j}{\| \hat{I}_j \| \| I_j^{gt} \|}.
    \label{eq:loss_color}
\end{equation} 

\noindent \textbf{Overall Spatial Loss:} The overall loss is a weighted combination of the aforementioned losses as follows:
\begin{equation}
    \mathcal{L}_s = \lambda_1 \mathcal{L}_r + \lambda_2 \mathcal{L}_{sm} + \lambda_3 \mathcal{L}_c.
    \label{eq:loss_spatial}
\end{equation}

\begin{table*}[!ht]
\centering
\setlength{\tabcolsep}{0.5pt}
\begin{tabular}{ccccc} 
\includegraphics[width=0.2\textwidth]{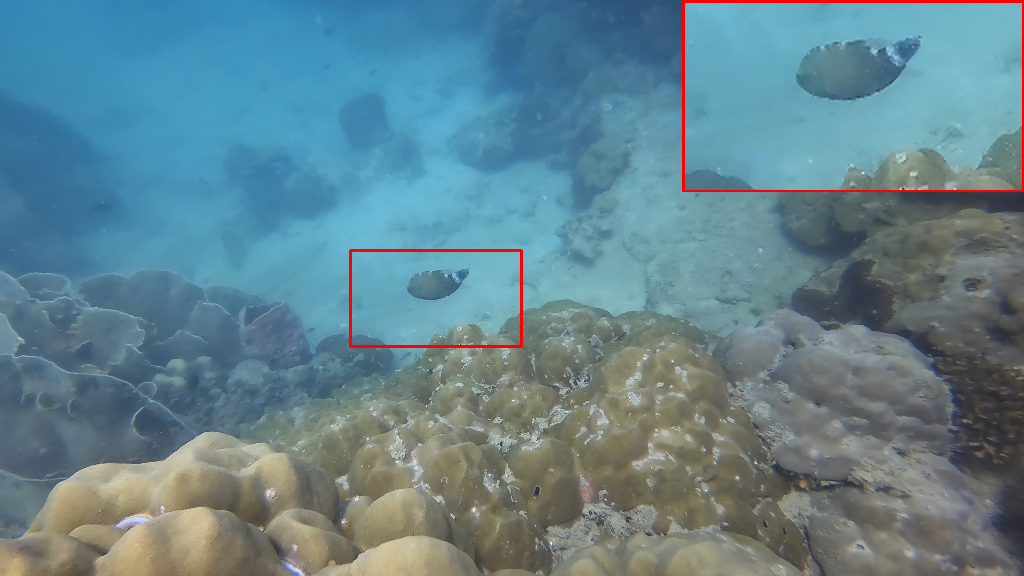}     &    \includegraphics[width=0.2\textwidth]{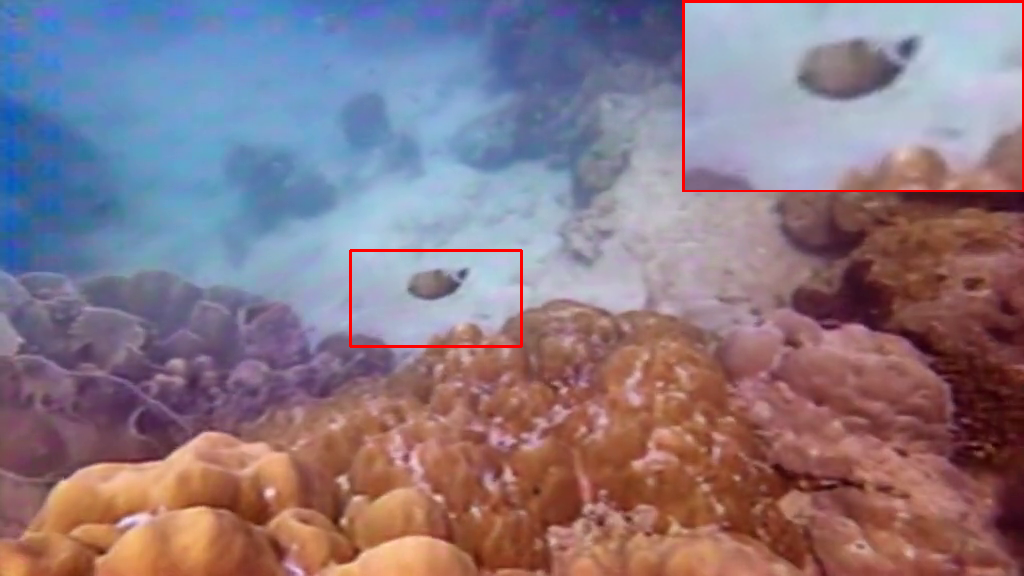} & \includegraphics[width=0.2\textwidth]{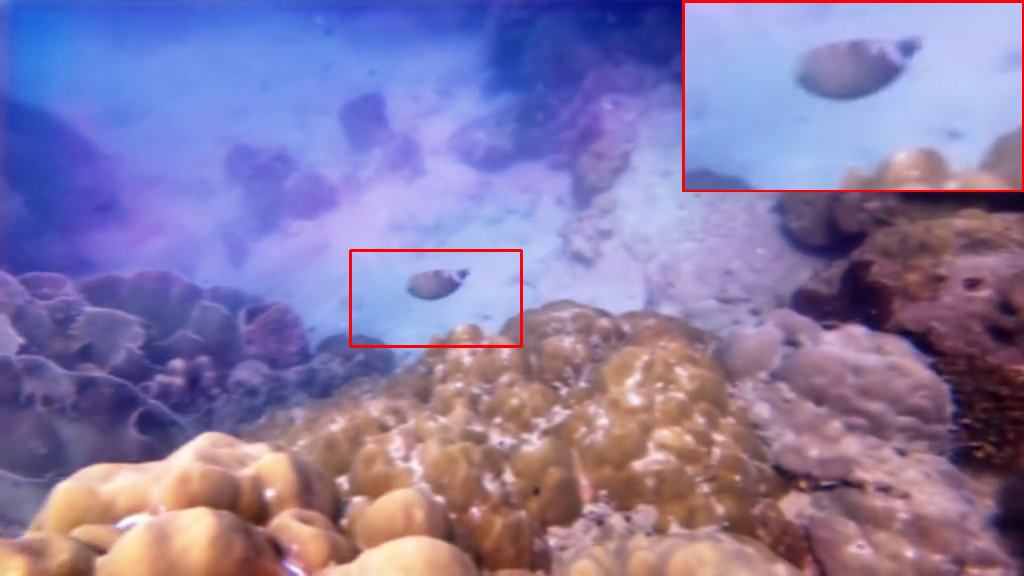}     & 
\includegraphics[width=0.2\textwidth]{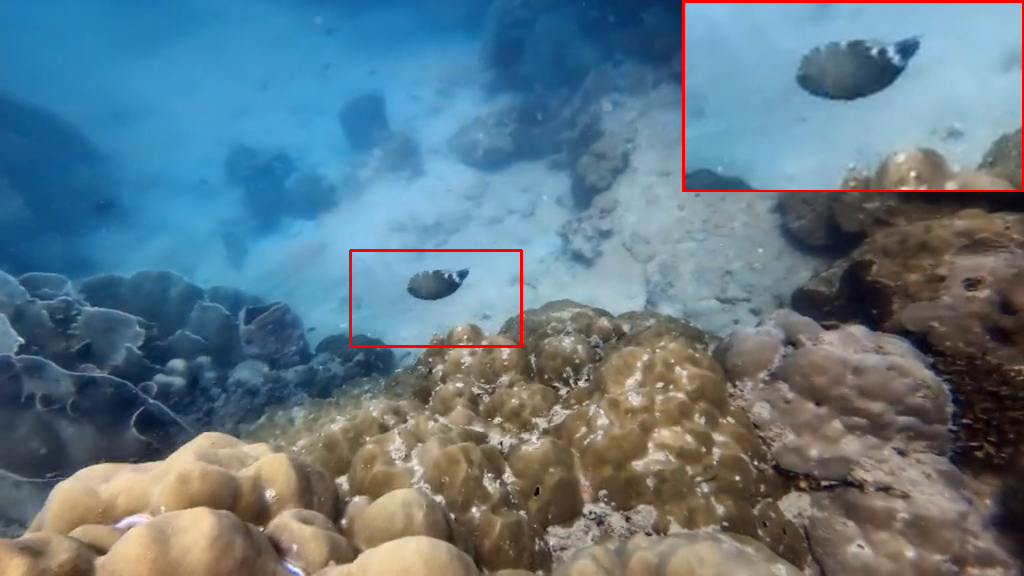}  &   \includegraphics[width=0.2\textwidth]{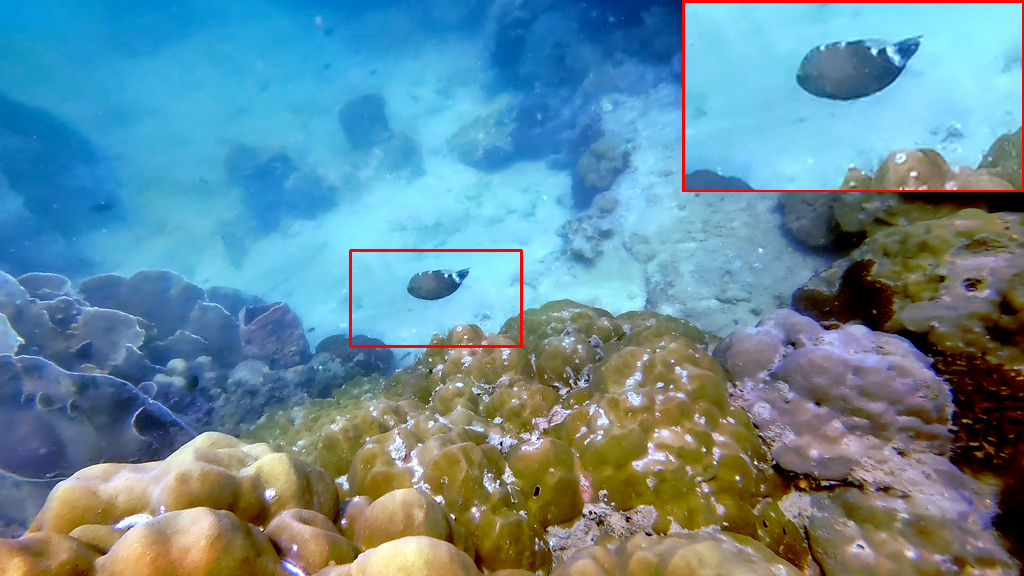}  \\
Raw Input & FunIE-GAN \cite{funiegan} & UW-GAN \cite{uwgan} & TOPAL \cite{topal} & PhISH-Net \cite{PhISHnet} \vspace{1mm}\\  
\includegraphics[width=0.2\textwidth]{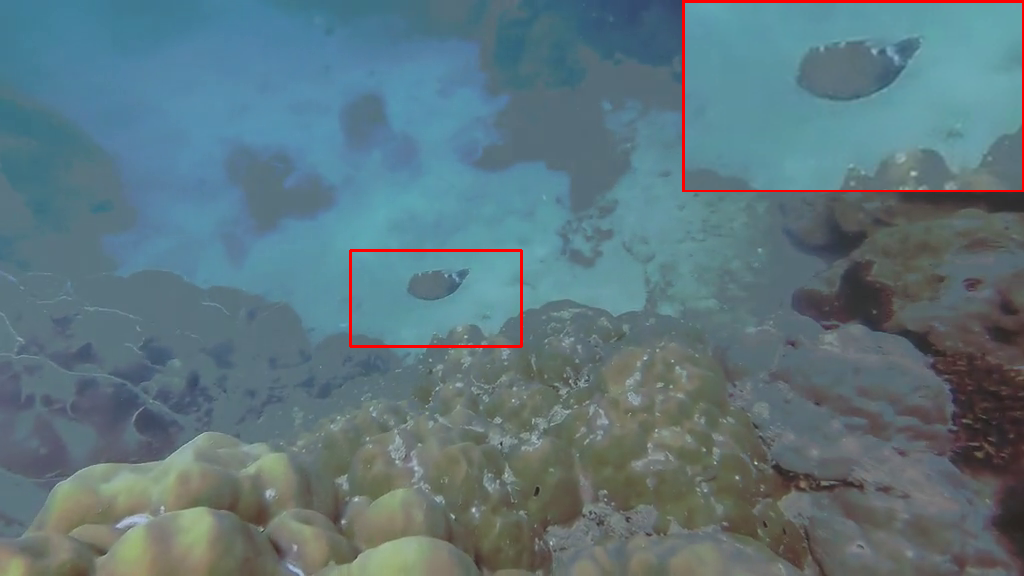}     &    \includegraphics[width=0.2\textwidth]{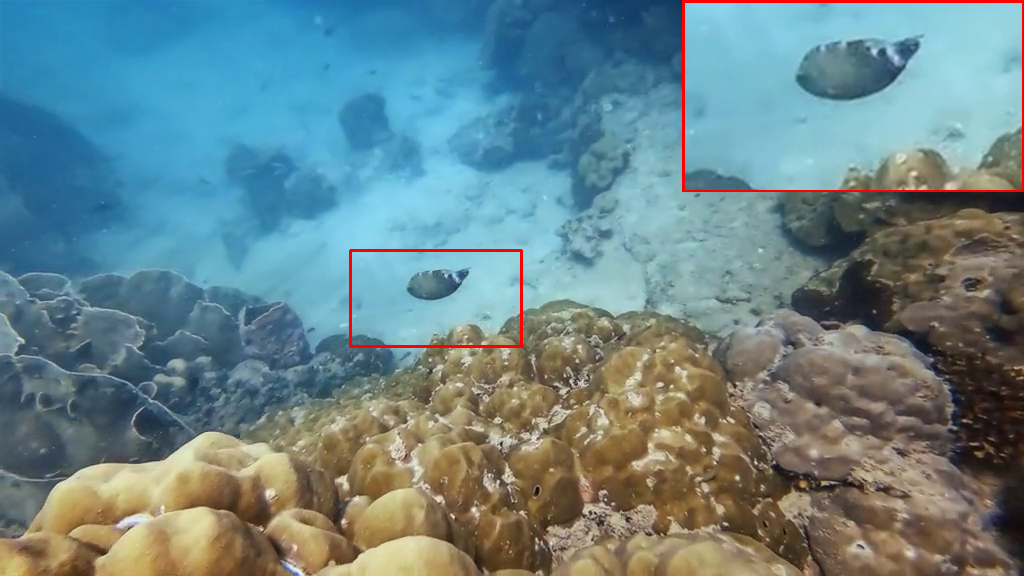}     & 
\includegraphics[width=0.2\textwidth]{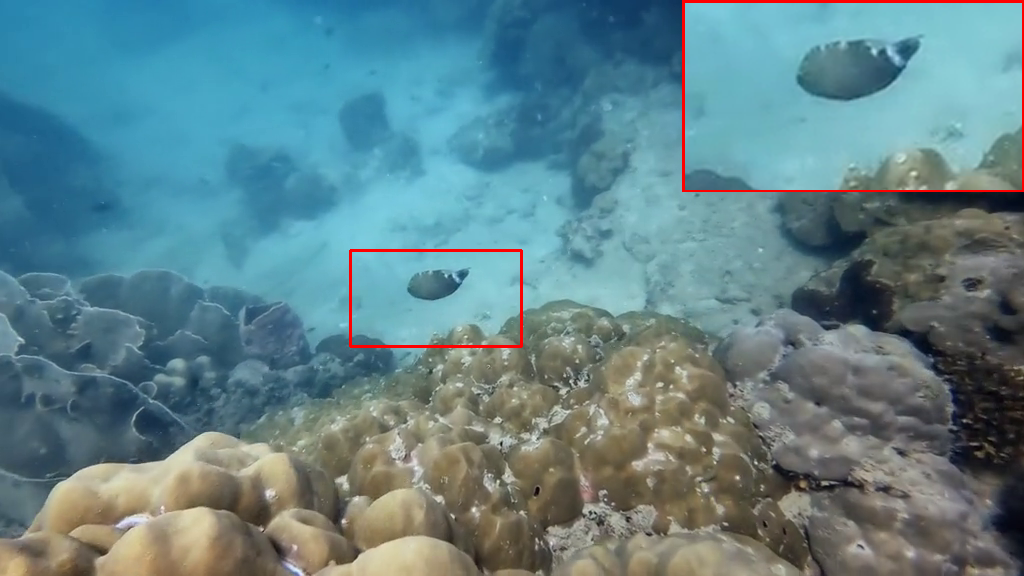}  &   \includegraphics[width=0.2\textwidth]{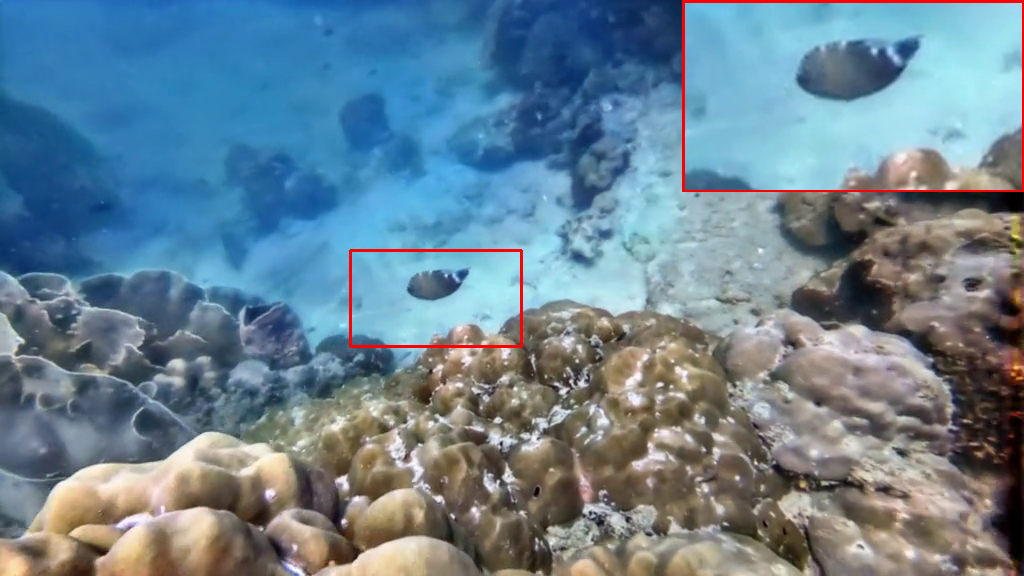}     &  \includegraphics[width=0.2\textwidth]{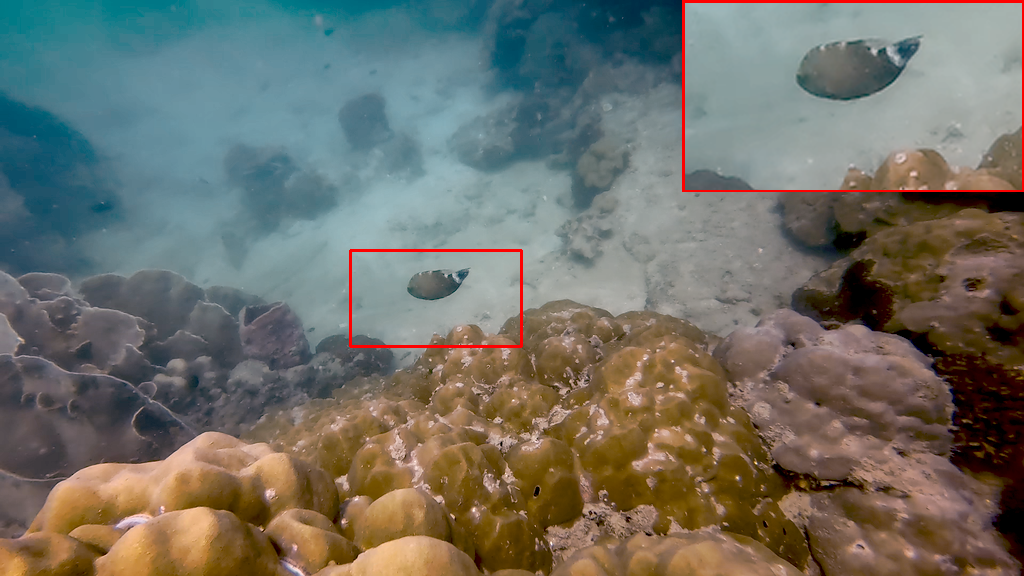} \\
USUIR \cite{usuir} & URanker \cite{uranker} & PUIE-Net \cite{puienet} & CLUIE-Net \cite{cluienet} & UnDIVE (Ours) \\ 
\end{tabular}
\captionof{figure}{Results of different enhancement methods on frame $54$ of the PhuQuoc1\_Jun2022.mp4 video from the MVK \cite{mvk} dataset. The blue hue in the scene is efficiently reduced by UnDIVE, while also improving the contrast in the enhanced image.}
\label{fig:comp_images}
\end{table*}

\subsection{Learning Temporal Consistency}

Image-based methods for underwater video enhancement (UVE) typically aggregate individually enhanced frames into a single video. However, this approach often results in temporal artifacts such as flickering, uneven illumination, and stabilization issues.
We propose to learn temporally consistent enhanced frames through a loss based on optical flow \cite{hornschunkof}.
Consider a pair of spatially enhanced frames $\hat{x}_t$ and $\hat{x}_{t+1}$.
The optical flow $\mathbf{U}_{t,t+1}$ is a dense estimate of vectors describing the motion of pixels from frame $\hat{x}_t$ to the adjacent frame $\hat{x}_{t+1}$.
This flow map can then be used to construct $\hat{x}_{t+1}$ from $\hat{x}_{t}$ via a warping operation $\mathcal{W}$ described as
\begin{equation}
    \hat{x}_t(\mathbf{p}) = \hat{x}_{t+1}(\mathbf{p} + \mathbf{U}_{t,t+1}(\mathbf{p})) = \mathcal{W}(\hat{x}_{t+1},\mathbf{U}_{t,t+1}),
    \label{eq:optical_flow}
\end{equation}
where $\mathbf{p}$ denotes the pixel coordinates. 
Let the spatial enhancement model be denoted by $f_\Theta$ (parameterized by $\Theta$).
The optical flow-based loss is optimized in an unsupervised manner given by 
\begin{align}
    \nonumber \mathcal{L}_{of}(x_t, x_{t+1}) = & \| \mathcal{W}(f_\Theta(x_{t}), \mathbf{U}_{t+1,t}) \\
    & - f_\Theta(\mathcal{W}(x_t, \mathbf{U}_{t+1,t})) \|_2^2.
    \label{eq:loss_of}
\end{align}
We consider the above loss in both forward and backward directions and utilize the total consistency loss $\mathcal{L}_t$ as
\begin{equation}
    \mathcal{L}_t  = 0.5 \mathcal{L}_{of}(x_t, x_{t+1}) + 0.5 \mathcal{L}_{of}(x_{t+1}, x_t).
    \label{eq:loss_temporal}
\end{equation}
This ensures that the optical flow warps the outputs from $f_\Theta$ in the same way that $f_\Theta$ enhances the warped frames.
Moreover, it also introduces temporal stability in the reconstructions since the L$_2$ loss penalizes large motion outliers resulting in high errors in intensities between the pixels. 
We obtain the optical flow maps for video frame pairs using an off-the-shelf FastFlowNet \cite{fastflownet}, which provides real-time and efficient flow estimates.\\

\noindent \textbf{Training UnDIVE:} After the generative prior is learned, we train UnDIVE first using $N_s$ epochs on underwater images using Eqn. \eqref{eq:loss_spatial} followed by $N_t$ epochs of training with temporal consistency in Eqn. \eqref{eq:loss_temporal}. 

\begin{table*}[!ht]
\adjustbox{max width=\textwidth}
\centering
\resizebox{\textwidth}{!}{%
\begin{tabular}{clccccccccc}
\toprule
 & \multicolumn{1}{c}{} & \multicolumn{6}{c}{Image Quality Metrics} & \multicolumn{3}{c}{ Video Quality Metrics} \\
 \cmidrule(lr){3-8}
 \cmidrule(lr){9-11}
\multirow{-2}{*}{ Dataset} & \multicolumn{1}{c}{\multirow{-2}{*}{Method}} & UCIQE &  UIQM &  UIConM &  UISM &  UICM &  CCF &  VSFA &  FastVQA &  DOVER \\
& & ($\uparrow$) & ($\uparrow$) & ($\uparrow$) & ($\uparrow$) & ($\uparrow$) & ($\uparrow$) & ($\uparrow$) & ($\uparrow$) & ($\uparrow$)\\
\midrule
 & FunIE-GAN \hfill \cite{funiegan} & 0.5412 & 0.7228 & 0.5774 & 2.1575 & 4.1292 & 14.7077 & 0.6212 & 0.3252 & 2.2951 \\
 & UW-GAN \hfill \cite{uwgan} & 0.5340 & 0.6758 & 0.5607 & 1.8062 & 3.4402 & 11.9872 & 0.6330 & 0.4720 & 4.3301 \\
 & TOPAL \hfill \cite{topal} & 0.5282 & 0.7325 & 0.5997 & 2.1860 & 2.3578 & 15.7300 & 0.7027 & 0.6588 & 7.2169 \\
 & PhISH-Net \hfill \cite{PhISHnet} & \textcolor{purple}{0.6256} & \textcolor{purple}{1.2666} & \textcolor{purple}{0.8945} & \textcolor{blue}{5.0350} & \textcolor{purple}{8.9778} & \textcolor{purple}{52.6171} & \textcolor{blue}{0.7395} & 0.5520 & 5.2110 \\
 & USUIR \hfill \cite{usuir} & 0.4806 & 0.5095 & 0.4242 & 1.5376 & 0.5568 & 6.6373 & 0.6169 & 0.4029 & 4.0153 \\
 & URanker \hfill \cite{uranker} & 0.5639 & 0.9126 & 0.7507 & 2.5199 & 4.6189 & 14.0485 & 0.7323 & 0.6711 & 7.2502\\
 & PUIE-Net \hfill \cite{puienet} & 0.5281 & 0.8216 & 0.6677 & 2.4717 & 3.0629 & 12.9824 & 0.7306 & \textcolor{blue}{0.6914} & \textcolor{blue}{7.7974} \\
\multirow{-8}{*}{\begin{turn}{90} VDD-C \cite{vddc} \end{turn}} & CLUIE-Net \hfill \cite{cluienet} & 0.5209 & 0.8431 & 0.6867 & 2.5228 & 3.0920 & 16.9909 & 0.6946 & 0.3508 & 6.9858 \\
\midrule
 & \textbf{UnDIVE} & \textcolor{blue}{0.5768} & \textcolor{blue}{1.2222} & \textcolor{blue}{0.8659} & \textcolor{purple}{5.2803} & \textcolor{blue}{5.8755} & \textcolor{blue}{47.1940} & \textcolor{purple}{0.7490} & \textcolor{purple}{0.6989} & \textcolor{purple}{7.9712} \\
\bottomrule
\vspace{-2.5mm}\\

 & FunIE-GAN \hfill \cite{funiegan} & \textcolor{purple}{0.5978} & 0.6084 & 0.5189 & 1.4411 & \textcolor{purple}{3.2392} & 14.8027 & 0.5126 & 0.2920 & 3.1907 \\
 & UW-GAN \hfill \cite{uwgan} & \textcolor{blue}{0.5857} & 0.4960 & 0.4331 & 1.0128 & \textcolor{blue}{3.0605} & 9.3128 & 0.5677 & 0.4397 & 3.6762 \\
 & TOPAL \hfill \cite{topal} & 0.5584 & 0.6888 & \textcolor{blue}{0.5919} & \textcolor{purple}{1.7384} & 1.9933 & \textcolor{purple}{28.5244} & 0.5850 & 0.3545 & 3.1957 \\
 & PhISH-Net \hfill \cite{PhISHnet} & 0.5480 & 0.5056 & 0.4515 & 0.9529 & 2.6692 & 13.5727 & 0.6421 & 0.3932 & 3.0815 \\
 & USUIR \hfill \cite{usuir} & 0.5414 & \textcolor{blue}{0.6724} & 0.5798 & \textcolor{blue}{1.6155} & 2.5423 & 7.2011 & \textcolor{blue}{0.6495} & 0.4235 & \textcolor{purple}{4.8267} \\
 & URanker \hfill \cite{uranker} & 0.5489 & 0.5754 & 0.5369 & 0.8297 & 2.7944 & 8.7077 & 0.6366 & 0.2620 & 2.7925 \\
 & PUIE-Net \hfill \cite{puienet} & 0.5450 & 0.5039 & 0.4714 & 0.7404 & 2.1551 & 11.1319 & 0.6280 & \textcolor{blue}{0.4754} & 3.5308 \\
\multirow{-8}{*}{\begin{turn}{90} Brackish \cite{brackish} \end{turn}} &  CLUIE-Net \hfill \cite{cluienet} & 0.5766 & \textcolor{purple}{0.7480} & \textcolor{purple}{0.6778} & 1.3819 & 3.0139 & \textcolor{blue}{16.8279} & 0.6275 & 0.3508 & 3.2155 \\
\midrule
 & \textbf{UnDIVE} & 0.5734 & 0.5466 & 0.5086 & 0.9128 & 2.6213 & 14.4568 & \textcolor{purple}{0.6516} & \textcolor{purple}{0.5151} & \textcolor{blue}{3.9462} \\
\bottomrule
\vspace{-2.5mm}\\

 & FunIE-GAN \hfill \cite{funiegan} & 0.5584 & 1.0009 & 0.7291 & 3.8875 & 5.2395 & 24.5786 & 0.6145 & 0.1928 & 2.2915 \\
 & UW-GAN \hfill \cite{uwgan} & 0.5476 & 0.7475 & 0.6028 & 2.1883 & 4.0130 & 12.6571 & 0.6285 & 0.3116 & 3.3599 \\
 & TOPAL \hfill \cite{topal} & 0.5627 & 0.8673 & 0.7087 & 2.4709 & 4.1826 & 22.7635 & 0.6843 & 0.5345 & 6.0123 \\
 & PhISH-Net \hfill \cite{PhISHnet} & \textcolor{purple}{0.6068} & \textcolor{purple}{1.0523} & 0.7400 & \textcolor{purple}{4.2980} & \textcolor{purple}{6.6543} & \textcolor{purple}{40.1159} & 0.7051 & 0.4841 & 5.0983 \\
 & USUIR \hfill \cite{usuir} & 0.4950 & 0.8299 & 0.6512 & 2.8675 & 2.1537 & 9.1856 & 0.6742 & 0.5620 & 5.9473 \\
 & URanker \hfill \cite{uranker} & 0.5642 & 0.8622 & 0.7034 & 2.4719 & 4.6904 & 16.0681 & 0.6952 & \textcolor{purple}{0.6711} & 5.7439 \\
 & PUIE-Net \hfill \cite{puienet} & 0.5540 & 0.7422 & 0.6039 & 2.1370 & 3.6716 & 13.9865 & \textcolor{blue}{0.7074} & 0.5731 & \textcolor{blue}{6.4301} \\
\multirow{-8}{*}{\begin{turn}{90} UOT32 \cite{uot32} \end{turn}} &  CLUIE-Net \hfill \cite{cluienet} & 0.5609 & 0.9919 & \textcolor{purple}{0.7627} & 3.4334 & 4.4822 & 22.9237 & 0.6946 & 0.6052 & \textcolor{purple}{6.5477} \\
\midrule
 & \textbf{UnDIVE} & \textcolor{blue}{0.5870} & \textcolor{blue}{1.0454} & \textcolor{blue}{0.7585} & \textcolor{blue}{4.2328} & \textcolor{blue}{5.4106} & \textcolor{blue}{39.3131} & \textcolor{purple}{0.7165} & \textcolor{blue}{0.5782} & 6.3157 \\
\bottomrule
\vspace{-2.5mm}\\

 & FunIE-GAN \hfill \cite{funiegan} & 0.5857 & 0.9507 & 0.8204 & 1.9898 & 6.5807 & 15.8532 & 0.6503 & 0.3320 & 4.2376 \\
 & UW-GAN \hfill \cite{uwgan} & 0.5841 & 0.9470 & 0.8296 & 1.8691 & 6.1803 & 15.9746 & 0.6332 & 0.4973 & 5.2086 \\
 & TOPAL \hfill \cite{topal} & 0.5919 & 1.1485 & 0.9792 & 2.7174 & 6.1778 & 18.5322 & 0.7503 & 0.7500 & 8.1786 \\
 & PhISH-Net \hfill \cite{PhISHnet} & \textcolor{purple}{0.9528} & \textcolor{purple}{1.4684} & \textcolor{blue}{1.0858} & \textcolor{blue}{5.2185} & \textcolor{purple}{10.7098} & \textcolor{purple}{59.5795} & 0.7547 & 0.7486 & 8.2122 \\
 & USUIR \hfill \cite{usuir} & 0.5171 & 0.7412 & 0.6633 & 1.5070 & 2.6042 & 8.7099 & 0.7127 & 0.6671 & 6.1041 \\
 & URanker \hfill \cite{uranker} & 0.5973 & 1.2178 & 1.0197 & 3.0411 & 7.2402 & 18.2012 & \textcolor{purple}{0.7871} & \textcolor{blue}{0.7805} & \textcolor{blue}{8.5637} \\
 & PUIE-Net \hfill \cite{puienet} & 0.5776 & 1.0748 & 0.9244 & 2.4490 & 5.7571 & 17.0048 & \textcolor{blue}{0.7821} & 0.7655 & 8.3190 \\
\multirow{-8}{*}{\begin{turn}{90} MVK \cite{mvk} \end{turn}} &  CLUIE-Net \hfill \cite{cluienet} & 0.5909 & 1.2080 & 1.0319 & 2.8679 & 6.1420 & 26.7405 & 0.7319 & 0.7542 & 7.1492 \\
\midrule
 &  \textbf{UnDIVE} & \textcolor{blue}{0.6226} & \multicolumn{1}{c}{\textcolor{blue}{1.4537}} & \multicolumn{1}{c}{\textcolor{purple}{1.0864}} & \multicolumn{1}{c}{\textcolor{purple}{5.2944}} & \multicolumn{1}{c}{\textcolor{blue}{7.8019}} & \multicolumn{1}{l}{\textcolor{blue}{56.1459}} & \multicolumn{1}{c}{0.7689} & \multicolumn{1}{c}{\textcolor{purple}{0.7910}} & \multicolumn{1}{c}{\textcolor{purple}{9.0325}} \\
 \bottomrule
\end{tabular}}
\caption{Performance comparison of UnDIVE with other enhancement methods on four datasets: VDD-C \cite{vddc}, Brackish \cite{brackish}, UOT32 \cite{uot32} and MVK \cite{mvk} using various image and video quality metrics. The best and second best are highlighted with \textcolor{purple}{purple} and \textcolor{blue}{blue} respectively.}
\label{tab:main_table}
\end{table*}

\section{Experiments and Results}

\subsection{Training and Implementation Details}

All experiments were carried out on three NVidia Tesla v100 GPUs with 32GB VRAM each. 
For training the DDPM, we utilize about 100,000 carefully chosen random crops of size $256$ from the UIEB \cite{uieb} dataset and train the DDPM for 100 epochs with a learning rate of $10^{-4}$ and a batch size of $N_D = 24$.
To train the spatial enhancement, the network is first trained using $890$ paired images from UIEB using the spatial loss for $100$ epochs with a batch size of $N_s = 64$.
This is followed by $100$ epochs of training on $34,000$ frame pairs from the UVE-38k \cite{uve38k} dataset with both spatial and the unsupervised temporal consistency loss with a batch size of $N_t = 24$.
The loss weights in Eqn. \eqref{eq:loss_spatial} are chosen as $(\lambda_1, \lambda_2, \lambda_3) = (1.0, 0.2, 0.1)$. \\

\noindent \textbf{Datasets:} For evaluating the enhancements, we consider four underwater video datasets:
\textbf{VDD-C} \cite{vddc} - a dataset of images of divers, drawn from videos taken in pool and field environments, and primarily meant for diver detection and tracking.
\textbf{Brackish} \cite{brackish} - a dataset consisting of $9$ video sequences with multiple synthetic distortions. Brackish was created to study marine organisms in a brackish strait.
\textbf{UOT32} \cite{uot32} - 20 videos requiring enhancement from a dataset for benchmarking object tracking algorithms.
\textbf{MVK} \cite{mvk} - 10 videos from a large-scale high-resolution (UHD) dataset collected from seas around the world. \\

\noindent \textbf{Metrics:} We consider multiple image and video no-reference (NR) quality metrics.
The image metrics consist of \textbf{UCIQE} (Underwater Colour Image Quality Evaluation) \cite{uciqe}, \textbf{CCF} (Colorfulness Contrast Fog density index) \cite{ccf}, \textbf{UIQM} (Underwater Image Quality Measure), \textbf{UICM} (Underwater Image Colorfulness Measure), \textbf{UISM} (Underwater Image Sharpness Measure) and \textbf{UIConM} (Underwater Image Contrast Measure) \cite{uiqm_uiconm_uicm}.\\[-1.5ex]

\noindent
The video metrics consist of \textbf{VSFA} \cite{vsfa}, \textbf{FastVQA} (Fast Video Quality Assessment) \cite{fastvqa} and \textbf{DOVER} (Disentangled Objective Video Quality Evaluator) (the technical quality score of DOVER) \cite{dover}. All metrics indicate higher quality when their corresponding values are higher.\\[-1.5ex]

\subsection{Results and Comparisons}

\noindent \textbf{Comparing Methods:} We compare UnDIVE with multiple state-of-the-art enhancement methods: (1) \textbf{FunIE-GAN} \cite{funiegan} - a fast enhancement method based on GANs, (2) \textbf{UW-GAN} \cite{uwgan} - an unsupervised GAN-based method that performs dehazing, (3) \textbf{TOPAL} \cite{topal} - a perceptual adversarial fusion network for UIE, (4) \textbf{PhISH-Net} \cite{PhISHnet} - a physics-inspired UIE network that operates on an image formation model, (5) \textbf{USUIR} \cite{usuir} - an unsupervised underwater image restoration method using a homology constraint, (6) \textbf{URanker} \cite{uranker} - a ranking-based image quality assessment method that utilizes a histogram prior to learn UIE, (7) \textbf{PUIE-Net} \cite{puienet} - an uncertainty inspired UIE method that generates maximally probable enhancement outputs, and (8) \textbf{CLUIE-Net} \cite{cluienet} - a UIE model that learns from multiple available enhancement candidates.\\

\begin{table}[!t]
\centering
\setlength{\tabcolsep}{0.5pt}
\begin{tabular}{ccc} 
\includegraphics[width=0.16\textwidth]{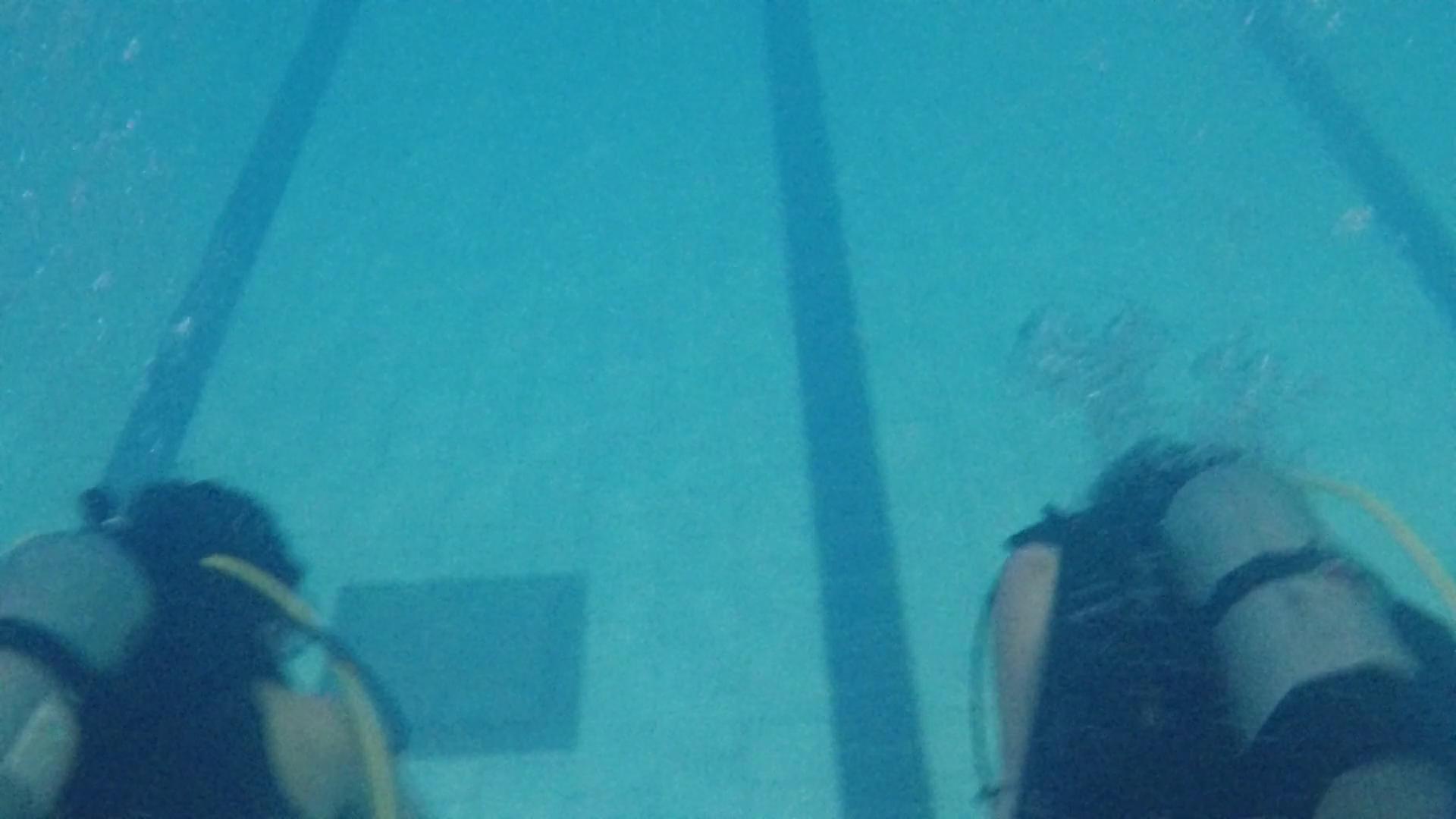}     &    \includegraphics[width=0.16\textwidth]{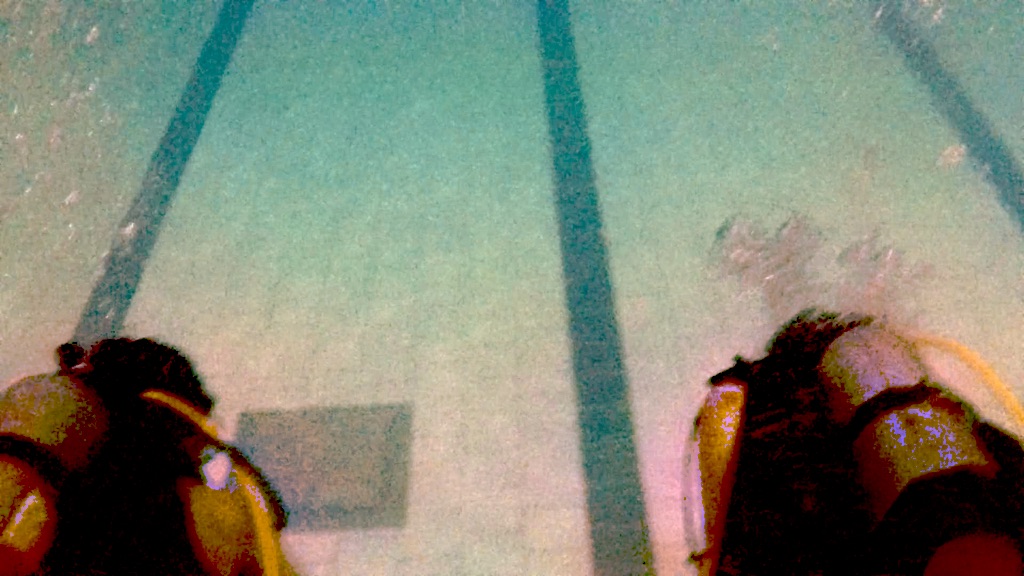} &  \includegraphics[width=0.16\textwidth]{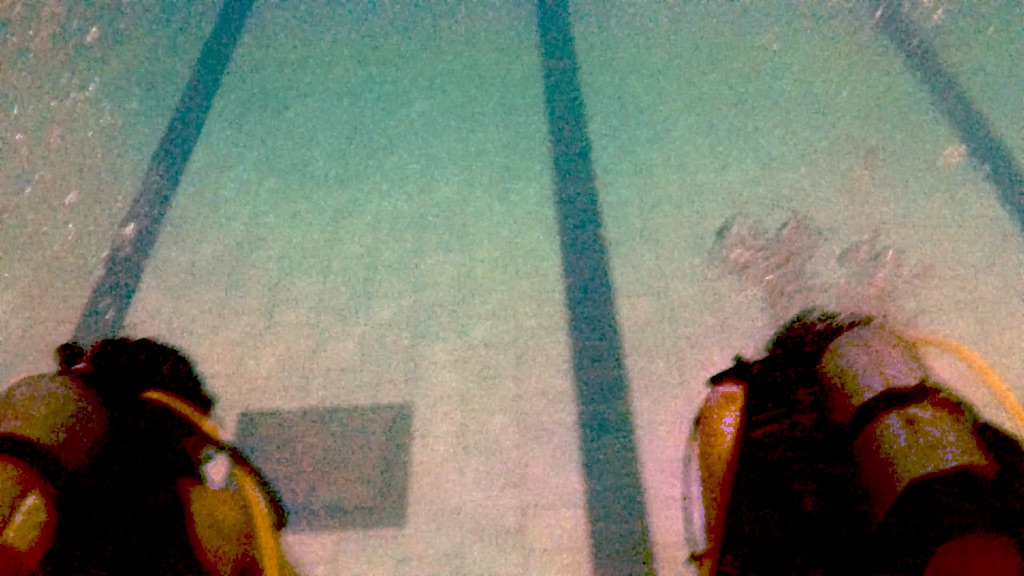}  \\
Raw Input & $\mathcal{L}_r$ & $\mathcal{L}_r + \mathcal{L}_{sm}$ \vspace{1mm}\\  
\includegraphics[width=0.16\textwidth]{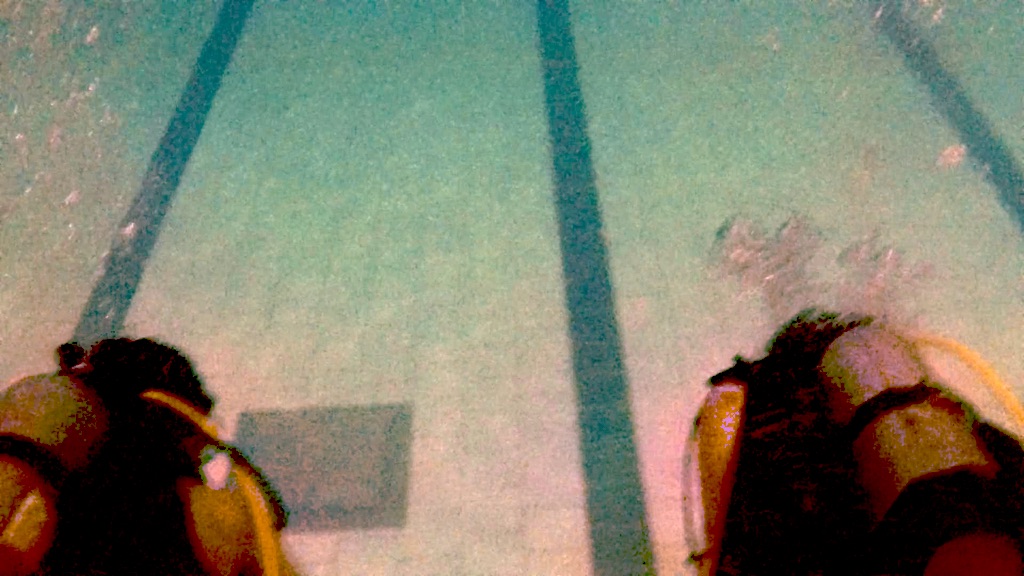}     &    \includegraphics[width=0.16\textwidth]{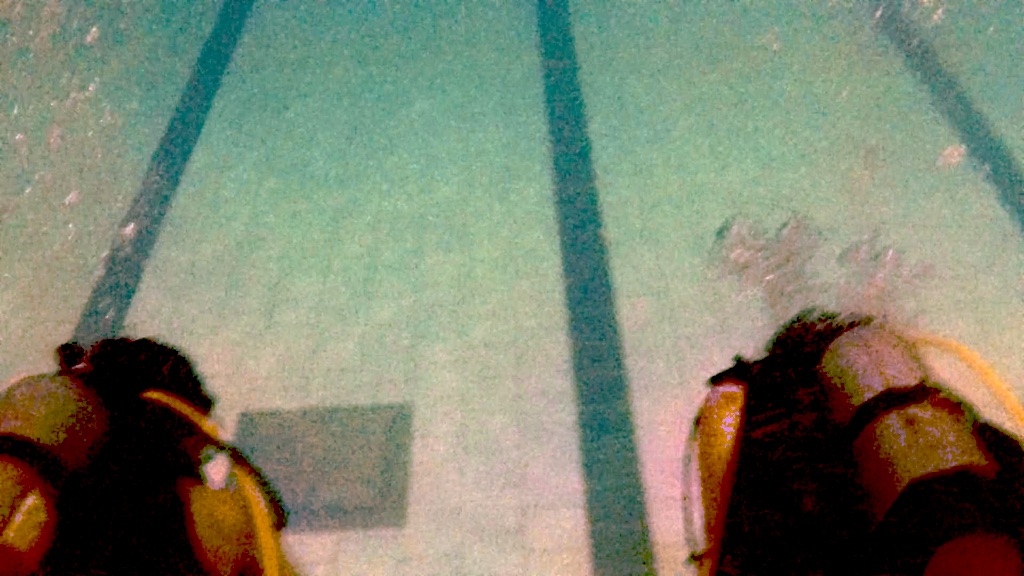}     & 
 \includegraphics[width=0.16\textwidth]{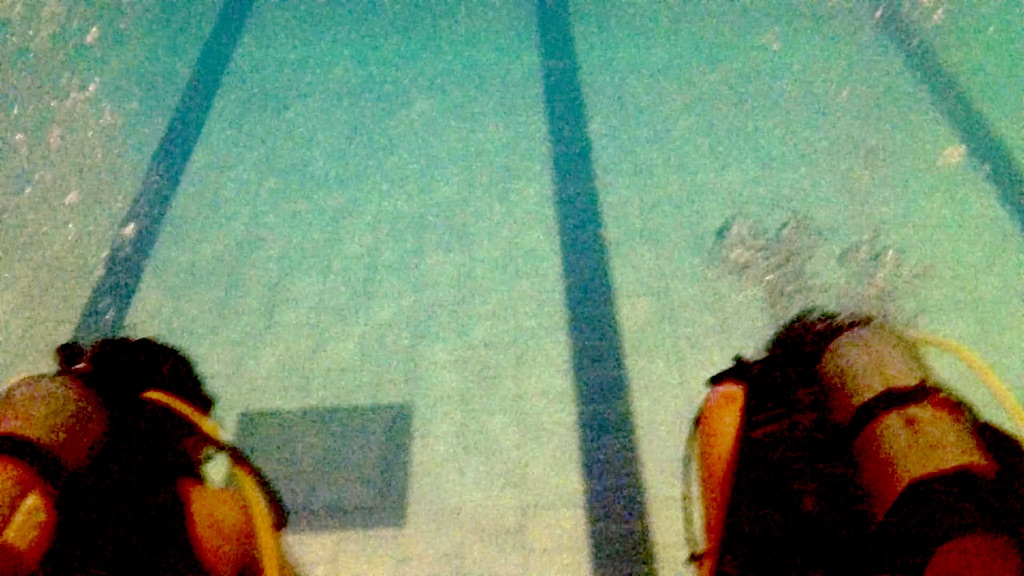} \\
 $\mathcal{L}_r + \mathcal{L}_c$ & $\mathcal{L}_r + \mathcal{L}_c + \mathcal{L}_{sm}$ & $\mathcal{L}_s + \mathcal{L}_t$ \\ 
\end{tabular}
\captionof{figure}{Results from the ablation study on the effect of different loss components. The model that uses $\mathcal{L}_s$ and $\mathcal{L}_t$ effectively removes the spurious reddish hue.}
\label{fig:ablation_losses}
\end{table}

\begin{table}[!t]
\adjustbox{max width=\columnwidth}
\centering
\begin{center}
\resizebox{0.8\columnwidth}{!}{%
\begin{tabular}{cccc|cc}
\toprule
$\mathcal{L}_r$ & $\mathcal{L}_{sm}$ & $\mathcal{L}_c$ & $\mathcal{L}_{t}$ & FastVQA & DOVER \\
\midrule
\checkmark &  &  &  & 0.6329 & 5.9192 \\
\checkmark & \checkmark &  &  & 0.6351 & 6.1339 \\
\checkmark &  & \checkmark &  & 0.6346 & 6.0749 \\
\checkmark & \checkmark & \checkmark &  & 0.6464 & 6.3021 \\ 
\checkmark & \checkmark & \checkmark & \checkmark & \textcolor{purple}{0.6989} & \textcolor{purple}{7.9712} \\
\bottomrule
\end{tabular}}%
\end{center}
\captionsetup{justification=centering}
\caption{Effect of different loss components of UnDIVE on UVE performance.}
\label{tab:ablation_losses}
\end{table}

\noindent \textbf{Quantitative Results:} We compare UnDIVE with the aforementioned methods and report all the quality metrics in Table \ref{tab:main_table}. 
To ensure fairness in all evaluations, we compute all metrics on enhanced videos with a fixed frame resolution of $1024 \times 576$.
We consistently see that UnDIVE outperforms other methods in terms of video metrics, but it is among the top two in image metrics. 
Since most of the UIE methods are trained on images, they naturally perform better on image metrics.
We note that UnDIVE performs well specifically on UISM, VSFA, FastVQA, and DOVER.
This could be attributed to better-restored sharpness in the enhanced frames along with reduced temporal artifacts like non-uniform illumination across frames. 
On VDD-C and MVK, UnDIVE consistently offers the best enhancement. However, for Brackish, image metrics show a drop and do not align with visual observations (detailed analysis in supplementary). For UOT32, while PhISH-Net performs best on image metrics, URanker and CLUIE-Net provide better temporal consistency. Overall, the metrics (video metrics in particular) confirm UnDIVE's strong generalization across diverse underwater scenes.\\

\noindent \textbf{Qualitative Evaluation:} Although the NR quality metrics provide a reasonable evaluation of different methods, they do not perfectly correlate with subjective human opinions.
Moreover, the lack of ground-truth data makes the quantitative evaluation challenging, making it important to observe the results from a visual perspective.
Methods apart from UnDIVE often generate color-based artifacts or reduced illumination in the enhanced images, as seen in Fig. \ref{fig:comp_images}.
UnDIVE significantly reduces water scattering effects and enhances colors, unlike other methods that fail to reduce the blue hue in the input.

\subsection{Ablation Experiments}

\noindent \textbf{Effect of Loss Components:} We analyze the effect of each component of the spatial loss in Eqn. \eqref{eq:loss_spatial} and the temporal loss in Eqn. \eqref{eq:loss_temporal}.
From Table \ref{tab:ablation_losses}, we observe that introducing each loss consistently improves the performance of the framework.
From the bottom two rows in Table \ref{tab:ablation_losses}, we observe that the temporal consistency loss gives a significant improvement in both video metrics, validating its efficacy in UVE. 
We also provide some visual enhancement results from using the loss components in training UnDIVE in Fig. \ref{fig:ablation_losses}.
We observe that using the reconstruction loss along with smoothness loss creates reddish coloration around the divers.
With the introduction of all the spatial loss components (collectively denoted by $\mathcal{L}_s$) along with the temporal loss $\mathcal{L}_t$ we observe the resolution of the color issue as well as a reduction in the blue color in the input image. \\

\begin{table}[!t]
\adjustbox{max width=\columnwidth}
\centering
\begin{center}
\resizebox{0.8\columnwidth}{!}{%
\begin{tabular}{cc|cc}
\toprule
DDPM Prior & Image Pretraining & FastVQA & DOVER \\
\midrule
 &  & 0.4830 & 4.7603 \\
 & \checkmark & 0.6876 & 7.7389 \\
\checkmark &  & 0.6678 & 7.0849 \\ 
\checkmark & \checkmark & \textcolor{purple}{0.6989} & \textcolor{purple}{7.9712} \\
\bottomrule
\end{tabular}} %
\end{center}
\captionsetup{justification=centering}
\caption{Effect of generative prior and pre-training on underwater image data on UVE performance.}
\label{tab:ablation_ddpm}
\end{table}

\noindent \textbf{Effect of Generative Prior and Image Pre-training:} We study the effectiveness of learning the generative prior and pre-training the spatial enhancement module on images from UIEB \cite{uieb}.
In cases where the prior is not used, the encoder $\mathcal{E}$ is trained from scratch along with the rest of the trainable parameters of $f_\Theta$.
Table \ref{tab:ablation_ddpm} shows that using the prior significantly improves performance compared to an encoder trained from scratch, indicating that learning robust image representations enhances the results. Additionally, pre-training on UIEB before training on video frame-pairs provides a substantial performance boost by leveraging diverse underwater scene information. Visual results in Fig. \ref{fig:ablation_ddpm} show that models trained without the prior or image pre-training reconstruct colors inaccurately. With only image pre-training, the enhanced image suffers from poor illumination, while with only the prior, it shows better contrast but a greenish hue.
When both components are used, the colors are more vivid and accurate.
More detailed analyses and experiments are provided in the supplementary. 

\begin{table*}[!t]
\centering
\setlength{\tabcolsep}{0.5pt}
\begin{tabular}{ccccc} 
\includegraphics[width=0.2\textwidth]{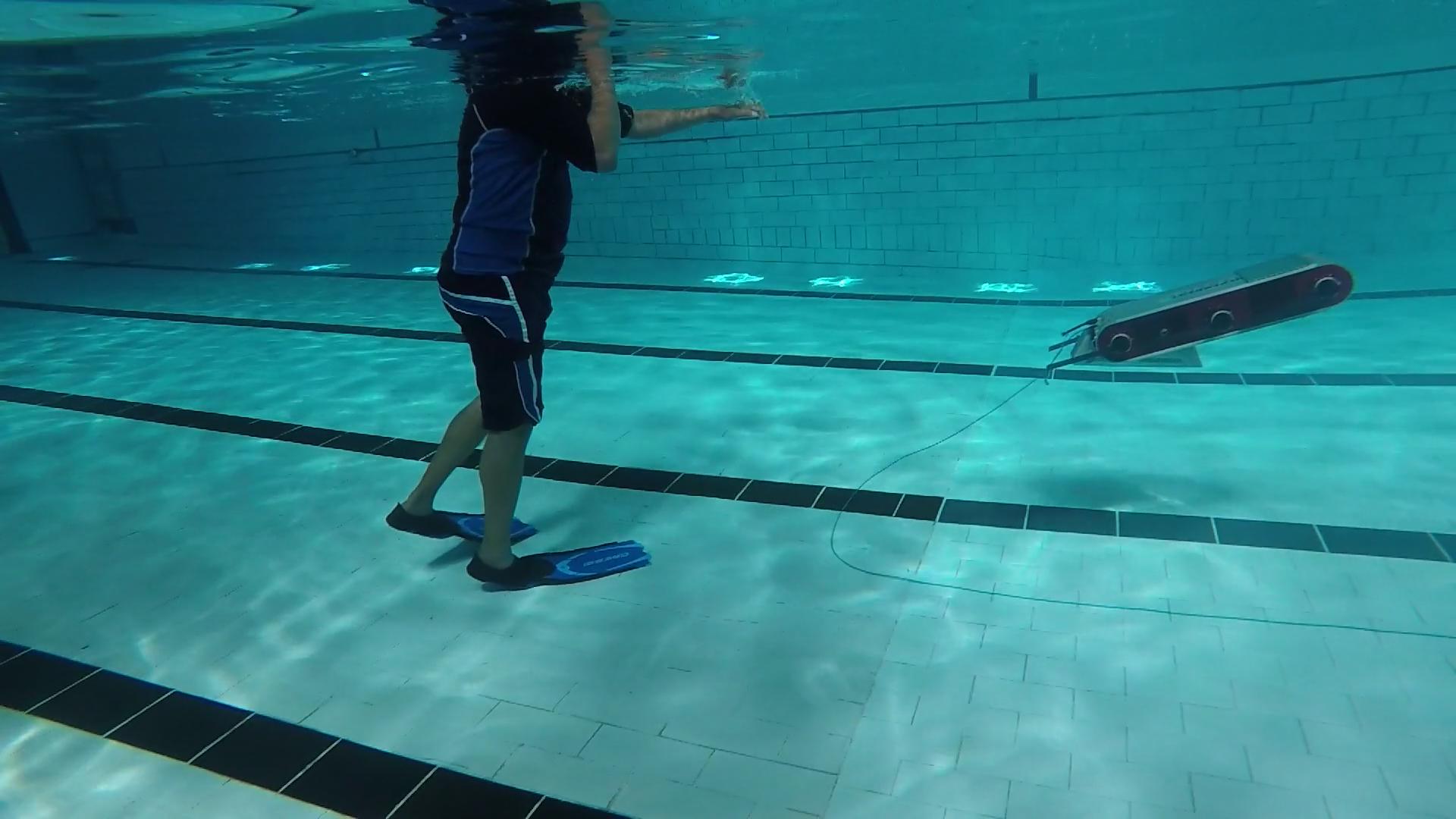}     &    \includegraphics[width=0.2\textwidth]{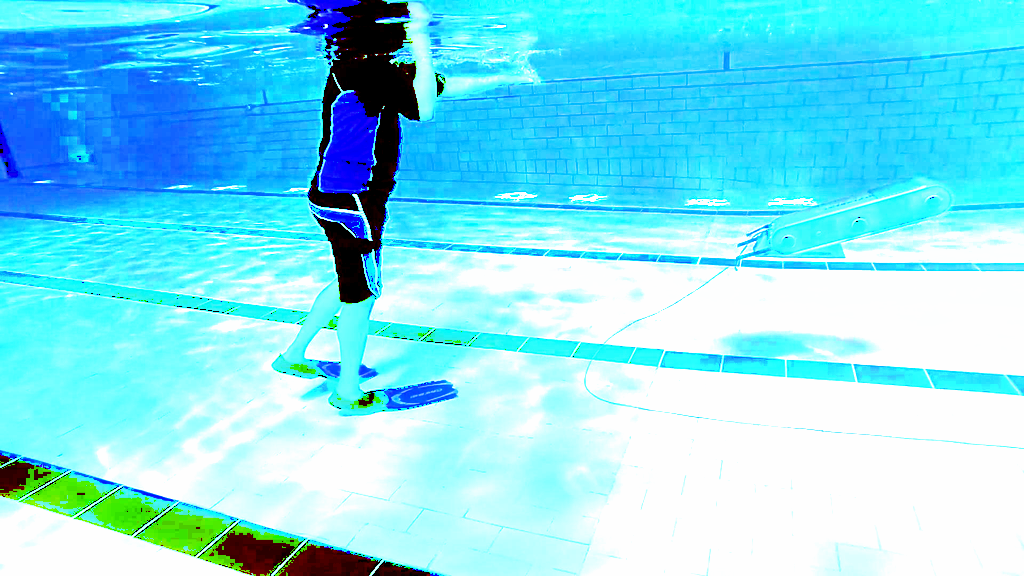} &  \includegraphics[width=0.2\textwidth]{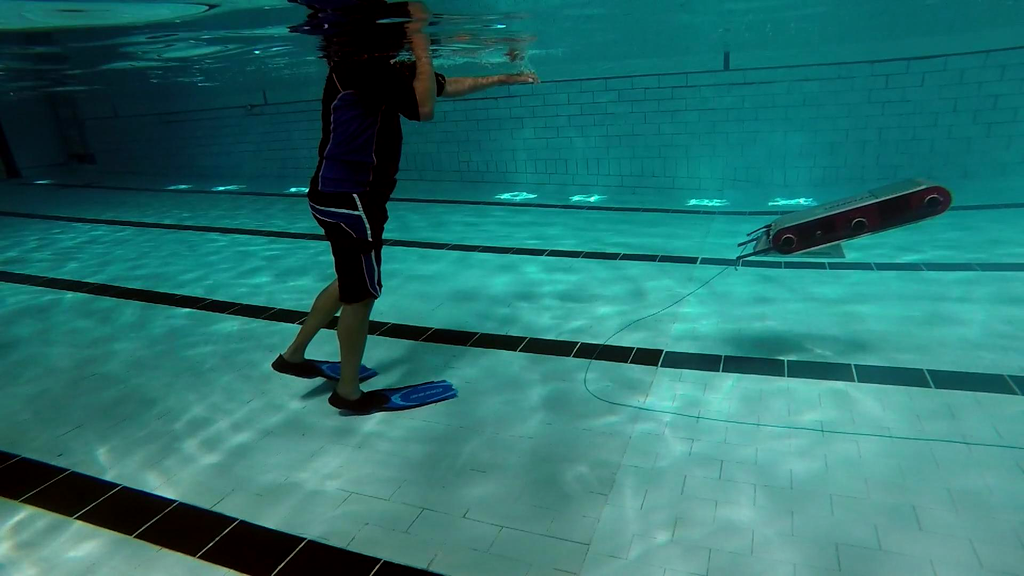}  
 & \includegraphics[width=0.2\textwidth]{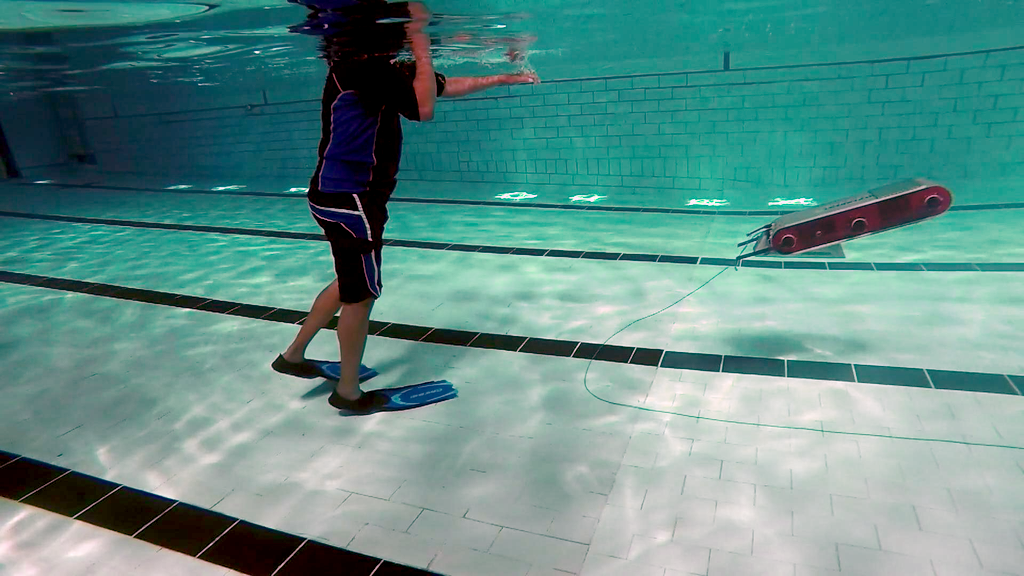}  & 
 \includegraphics[width=0.2\textwidth]{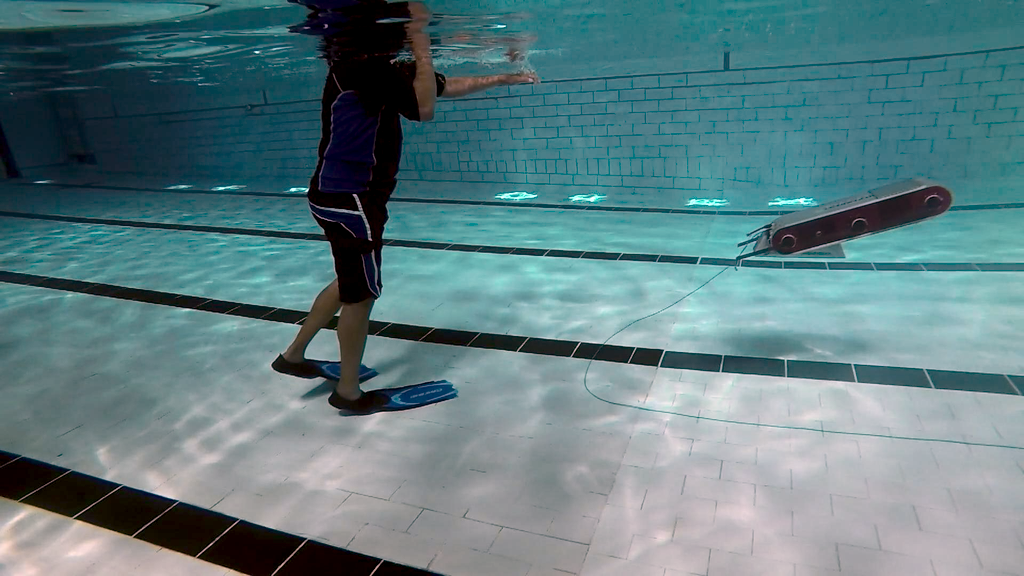} \\
 Raw Input & No prior or pre-training & Only pre-training & Only prior & Prior and pre-training \\ 
\end{tabular}
\captionof{figure}{Effect of the generative prior and the image pre-training on enhancement.}
\label{fig:ablation_ddpm}
\end{table*}

\begin{table}[!t]
\centering
\setlength{\tabcolsep}{0.5pt}
\begin{tabular}{ccc} 
\includegraphics[width=0.15\textwidth]{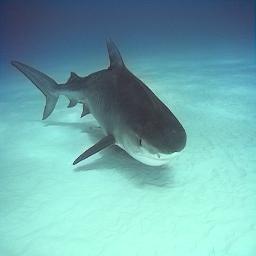}     &    \includegraphics[width=0.15\textwidth]{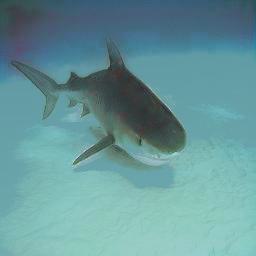} &  \includegraphics[width=0.15\textwidth]{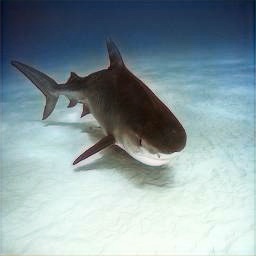}  \\
Raw Input & USUIR \cite{usuir} & URanker \cite{uranker} \vspace{1mm}\\  
\includegraphics[width=0.15\textwidth]{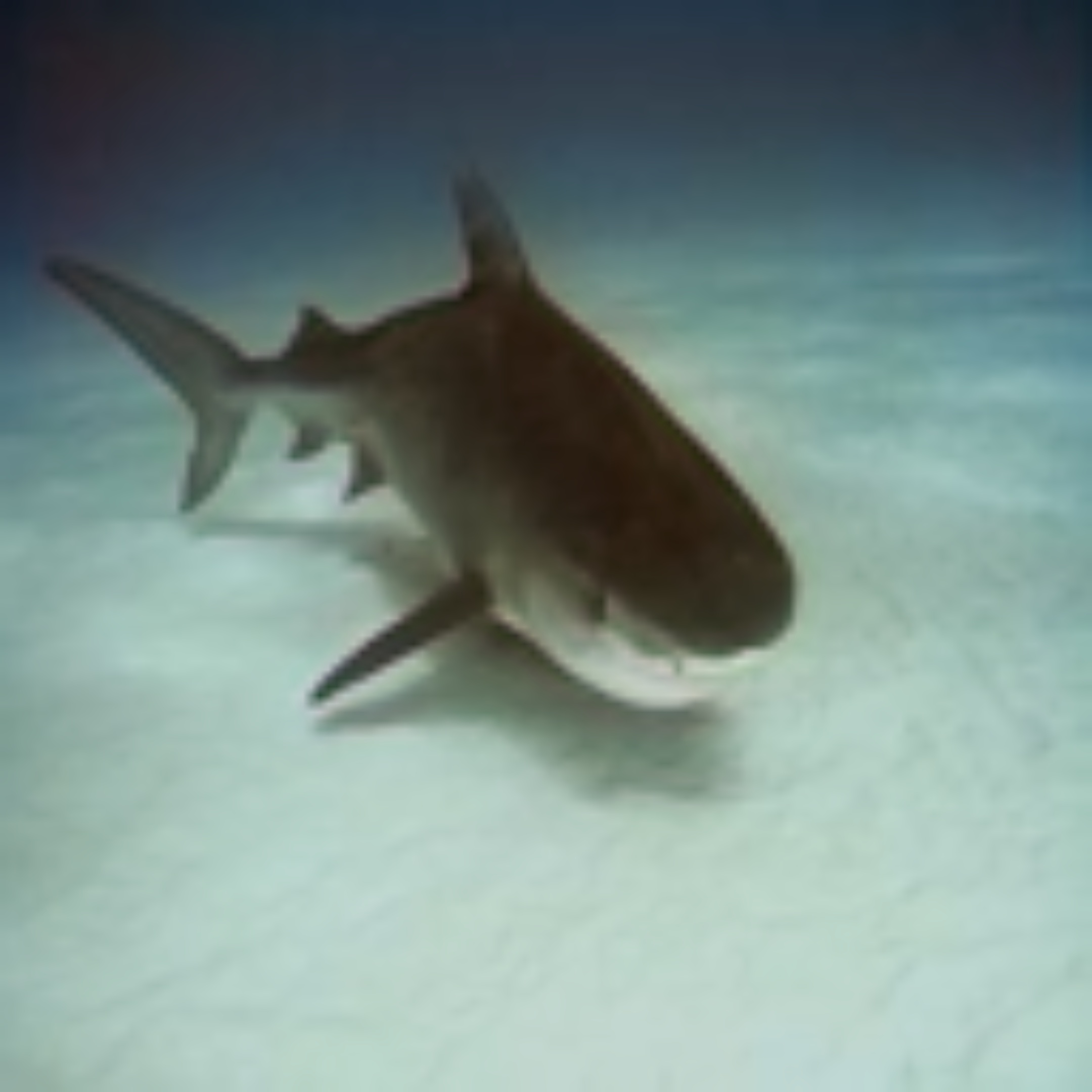}     &    \includegraphics[width=0.15\textwidth]{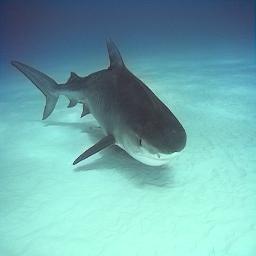}     & 
 \includegraphics[width=0.15\textwidth]{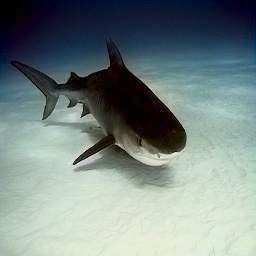} \\
 PUIE-Net \cite{puienet} & CLUIE-Net \cite{cluienet} & UnDIVE (Ours) \\ 
\end{tabular}
\captionof{figure}{Cross-dataset results of various UIE methods on EUVP.}
\label{fig:euvp_experiment}
\end{table}

\subsection{UnDIVE for Underwater Image Enhancement}

We enhance images from the EUVP \cite{euvp} and UIEB \cite{uieb} datasets in a cross-dataset setting and report the results in Table \ref{tab:uie_experiment}. 
We compare UnDIVE with four recent state-of-the-art UIE methods and notice that UnDIVE achieves best or second best performances consistently, validating its generalization capability beyond the task of UVE. 
From Fig. \ref{fig:euvp_experiment}, we notice that UnDIVE is able to efficiently enhance the input image.
URanker is able to provide better illumination while also reducing the blue hue.
However, other methods like USUIR and CLUIE-Net are unable to mitigate backscatter from the water scattering.

\subsection{Runtime Analysis}

Table \ref{tab:runtime} presents a runtime analysis, including model and computational complexity for all enhancement methods. While GAN-based methods FunIE-GAN and UW-GAN offer the fastest computation, they yield lower-quality enhancements. Per-frame computation times are shown for resolutions of $256 \times 256$ for GAN-based methods and $512 \times 512$ for the others.\\

\noindent \textbf{Limitations:} The lack of clear ground-truth references makes training with annotated frame-pairs challenging. Most quality assessment methods correlate poorly with human perception, making quantitative results somewhat misleading, particularly for image metrics. Video-based quality metrics, however, are more reliable for performance comparison (see supplementary for details).

\section{Conclusion}
\label{sec:conclusion}

\begin{table}[!t]
\adjustbox{max width=\columnwidth}
\centering
\begin{center}
\resizebox{\columnwidth}{!}{%
\begin{tabular}{l|ccc|ccc}
\hline
 & \multicolumn{3}{c|}{EUVP \cite{euvp}} & \multicolumn{3}{c}{UIEB \cite{uieb}} \\
\multirow{-2}{*}{Method} & PSNR ($\uparrow$) & SSIM ($\uparrow$) & UCIQE ($\uparrow$) & PSNR ($\uparrow$) & SSIM ($\uparrow$) & UCIQE ($\uparrow$) \\ \hline
USUIR & 14.4970 & 0.7093 & 0.5111 & 12.3075 & 0.6073 & 0.5432 \\
URanker & \textcolor{purple}{18.6961} & \textcolor{purple}{0.9315} & 0.5781 & \textcolor{blue}{20.4140} & \textcolor{purple}{0.9074} & \textcolor{blue}{0.6113} \\
PUIE-Net & 17.8718 & 0.7199 & 0.5720 & 19.4576 & 0.8560 & 0.5993 \\
CLUIE-Net & 17.8986 & 0.8886 & \textcolor{purple}{0.5895} & 13.7158 & 0.6833 & 0.6031 \\ \hline
UnDIVE & \textcolor{blue}{18.0893} & \textcolor{blue}{0.9002} & \textcolor{blue}{0.5884} & \textcolor{purple}{20.4598} & \textcolor{blue}{0.8717} & \textcolor{purple}{0.6167} \\ \hline
\end{tabular}}%
\end{center}
\caption{UIE performances on the EUVP (cross-dataset) and UIEB (intra-domain) datasets.}
\label{tab:uie_experiment}
\end{table}

\begin{table}[t]
\adjustbox{max width=\columnwidth}
\centering
\begin{center}
\resizebox{0.9\columnwidth}{!}{%
\begin{tabular}{lccc}
\toprule
Method & Runtime (s) ($\downarrow$) & GFLOPs ($\downarrow$) & Parameters (M) ($\downarrow$) \\
\midrule
FunIE-GAN & 0.00122 & 10.24 & 7.019\\
UW-GAN & 0.00037 & 0.004 & 1.925\\
TOPAL & 1.98956 & 111.6 & 36.67\\
PhISH-Net & 0.47244 & 0.090 & 0.556\\
USUIR & 0.06322 & 10.33 & 0.225\\
URanker & 0.03189 & 14.74 & 3.146\\
PUIE-Net & 0.21560 & 33.64 & 1.401\\
CLUIE-Net & 0.09292 & 24.68 & 13.39\\
\midrule
\textbf{UnDIVE} & 0.20908 & 7.153 & 6.723\\
\bottomrule
\end{tabular}}%
\end{center}
\caption{Runtime and Complexity Analysis of UnDIVE.}
\label{tab:runtime}
\end{table}

We propose a UVE method that effectively enhances underwater videos across various water types and degradations. 
This work is the first to use a generative prior from a self-supervised DDPM to guide the learning of a downstream enhancement task. 
By incorporating temporal consistency into the enhancement model, we demonstrate that UnDIVE can seamlessly generalize across multiple underwater video datasets (diverse water types) in real time. 
Furthermore, we show that UnDIVE's capabilities extend beyond UVE (for instance, UIE), potentially opening new avenues for research in other marine applications.

\subsection*{Acknowledgements}

\noindent Suhas Srinath acknowledges the Ministry of Education, India. 
Prathosh A. P. acknowledges support from Infosys foundation and IISc startup grant for computational resources.

{\small
\bibliographystyle{ieee_fullname}
\bibliography{egbib}
}

\end{document}